%% file: lism-sharp.tex
\documentclass[12pt]{article}

\usepackage{setspace}
\usepackage[margin=2.5cm]{geometry}

\usepackage{amsmath,amsthm,amssymb,dsfont, graphicx, tikz, makecell} 
\usepackage{caption}
\usepackage[margin=10pt,font=small,labelfont=bf,labelsep=period]{caption}

\usepackage{pgfplots} 
\usepackage{ifthen}
\usepackage{color,tikzsymbols,float,bm}

\usepackage{booktabs} 
\usepackage{multirow}
\usepackage{hhline}

\usepackage[absolute]{textpos}
\usepackage{paralist} 
\usepackage{appendix}

\usepackage{hyperref}
    \hypersetup{bookmarksnumbered=true, colorlinks=true,   
               citecolor=blue,urlcolor=blue,linkcolor=blue}

      \newcommand{\jimsecsref}[1]{\hyperref[#1]{Sections~\ref*{#1}}}

\usepackage{enumitem}
 
\usetikzlibrary{decorations.pathreplacing}
\usetikzlibrary{positioning}
\usetikzlibrary{patterns}
\usetikzlibrary{calc}
\usetikzlibrary{shapes}
\usetikzlibrary{math}
\usetikzlibrary{intersections}
\usetikzlibrary{fillbetween}
\usepgfplotslibrary{fillbetween}

\newtheorem{theorem}{Theorem}
\newtheorem{lemma}{Lemma}
\newtheorem{corollary}{Corollary}

\newtheorem{proposition}{Proposition}
\newtheorem{assumption}{Assumption}

\newcommand{\E}{\mathds{E}}
\newcommand{\F}{\mathcal{F}}
\newcommand{\U}{\mathcal{U}}

\newcommand{\A}{\mathcal{A}}

\newcommand{\M}{\mathcal{M}}

\newcommand{\R}{\mathcal{R}}
\newcommand{\W}{\mathcal{W}}
\newcommand{\I}{\mathds{I}}

\newcommand{\bp}{\overline{p}}
\newcommand{\bw}{\overline{w}}

\newcommand{\bx}{\bar{x}}

\newcommand{\mte}{\text{MTE}}

\newcommand{\late}{\text{LATE}}
\newcommand{\ate}{\text{ATE}}

\newcommand{\mprte}{\text{MPRTE}}
\newcommand{\te}{\text{WTE}}

\newcommand{\by}{\bar{y}}

\DeclareMathOperator*{\supp}{supp}

\makeatletter
\def\@biblabel#1{}  
\renewcommand\@openbib@code{%
      \advance\leftmargin\bibindent
      \itemindent -\bibindent
      \listparindent \itemindent
      \parsep \z@}
\def\@cite#1#2{({#1\if@tempswa , #2\fi})}  
\makeatother

\title{Sharp Bounds in the Latent Index Selection Model}
\author{Philip Marx%
   \thanks{
    Department of Economics, Louisiana State University, Baton Rouge, LA 70808. 
   Email: \href{mailto:philiplmarx@gmail.com}%
   {philiplmarx@gmail.com}.
   This paper is a revised and distilled version of my 2019 job market paper
   titled ``Within Compliers and Beyond: Sharp Bounds in Instrumental Variable
   Models.'' 
   I am especially grateful to Roland Fryer and Elie Tamer for helpful comments
   and guidance during this work.  
   This paper has also benefited from the helpful comments of Isaiah Andrews,
   Alex Bell, Ivan Canay, Carlos Flores, Magne Mogstad, Nicola Persico,
   Michael Powell, Ellie Prager, Alexander Torgovitsky, and
   participants in several seminars and conferences.
   The previous financial support of the Education Innovation Lab and the
   hospitality of the EO Foundation are gratefully acknowledged. 
   Of course, all errors are mine.  
}
}

\vspace{0.2in}

\date{\normalsize\today}

\begin{document}
\maketitle

\abstract{
    A fundamental question underlying the literature on partial identification
    is: what can we learn about parameters that are relevant for policy but not
    necessarily point-identified by the exogenous variation we observe?
    This paper provides an answer in terms of sharp, analytic characterizations
    and bounds for an important class of policy-relevant treatment effects,
    consisting of marginal treatment effects and linear functionals thereof, in
    the latent index selection model as formalized in Vytlacil~\cite{vytlacil}. 
    The sharp bounds use the full content of identified marginal distributions,
    and analytic derivations rely on the theory of stochastic orders. 
    The proposed methods also make it possible to sharply incorporate new
    auxiliary assumptions on distributions into the latent index selection
    framework. 
    Empirically, I apply the methods to study the effects of Medicaid on
    emergency room utilization in the Oregon Health Insurance Experiment, 
    showing that the predictions from extrapolations based on a distribution
    assumption (rank similarity) differ substantively and consistently from
    existing extrapolations based on a parametric mean assumption (linearity).     
    This underscores the value of utilizing the model's full empirical content
    in combination with auxiliary assumptions.
}

\singlespacing
\noindent
Keywords: instrumental variables, latent index selection model, marginal
treatment effects, partial identification, sharp bounds, counterfactuals,
extrapolation, second-order stochastic dominance, stochastic ordering,
majorization, Oregon Health Insurance Experiment\\[0.1in]
JEL Classification: C14, C26, C50

\onehalfspacing

\section{Introduction} 

The method of instrumental variables (IV) forms a cornerstone of modern
econometrics.  
In the basic program evaluation problem with imperfect compliance, Imbens and
Angrist~\cite{imbensangrist} show that IV yields an average causal effect of
treatment on outcomes when the instrument shifts treatment monotonically; 
this is the influential Local Average Treatment Effect ($\late$) on the subgroup
of compliers, whose treatment decisions depend on the instrument.     
By definition, the group of compliers --- and thus the identified causal effect
--- are inextricably tied to the instrument at hand. 

As a consequence, the variation that identifies a quantity of interest is not
always the variation that is observed.  
For example, typical quasi-experimental settings do not provide a choice of
instrument ---  the instrument may be too coarse, or it may not correspond to a
counterfactual of interest. 
Even when such a choice is available, as in the design of experiments, it is
often constrained in practice by other considerations such as statistical power
or limited resources. 
Furthermore, assuming that a treatment effect of interest is equal to the one
that is identified precludes selection on unobservables, which belies the
purpose of the model. 
This motivates an alternative structural approach to identification based on 
a marginal treatment effect ($\mte$) function in an empirically equivalent
latent index selection model [Heckman and Vytlacil~\cite{hv1999, hv2005};
Vytlacil~\cite{vytlacil}]. 
A fundamental question, raised recently within this paradigm by Mogstad,
Torgovitsky, and Santos~\cite{mogstad}, is the following: what can we learn
about parameters that are relevant for policy but not necessarily
point-identified by the exogenous variation we observe?%
\footnote{
    Of course, the question transcends this paradigm. 
    For previous sharp bounds on average treatment effects under exclusion
    restrictions, see Manski~\cite{manski1989, manski, manski94, manski03},
    Balke and Pearl~\cite{balkepearl}, and Heckman and Vytlacil~\cite{hv2000};
    for additional sharp bounds on the potential outcome distribution under
    varying exclusion restrictions, see Kitagawa~\cite{kitagawa09}; for sharp
    bounds on effects on the treated, see Huber et al.~\cite{huber}. 
    Other related work considers similar exclusion and monotonicity assumptions
    but derives bounds on different parameters or in different frameworks. 
    For respective examples, Tian and Pearl~\cite{tianpearl2000} derive bounds
    on probabilities of causation, and Zhang and Rubin~\cite{zhangrubin2003}
    derive bounds on treatment effects in a model with truncation by ``death,''
    where some potential outcomes are undefined in the traditional sense.
}
The question nests a variety of applications, including inference about
counterfactuals, inference about internally valid parameters that may not be
point-identified, and specification tests. 

This paper provides an answer in terms of sharp, analytically-derived bounds
for an important class of target parameters in the latent index selection model,
as formulated in Vytlacil~\cite{vytlacil}.   
Thus the paper builds most directly on the recent yet already influential
contribution of Mogstad et al.~\cite{mogstad}, who introduce a computational
framework for counterfactual inference that utilizes --- and exhausts --- the
identifying power of observed conditional means for a closely related class of
target parameters.
A starting point of the present contribution is that consistency with
conditional means --- henceforth, mean consistency --- does not exhaust the
identifying power of the fully independent latent index selection model, which
is frequently applicable in settings with (quasi)random assignment.   
In the absence of further assumptions or structure, mean consistency only
provides information about point-identified mean outcomes 
and the $\late$ among compliers.  
In addition to means, the empirically equivalent $\late$ model identifies
distributions of potential outcomes among certain subpopulations [Imbens and
Rubin~\cite{imbensrubin}, Abadie~\cite{abadie}].  
Consistency with identified distributions --- henceforth, distribution
consistency --- 
imposes additional restrictions on target parameters, yielding some information
about parameters other than the observed means and the $\late$, even in the
basic model. 

As a motivating example, suppose an experiment subsidized treatment relative to
a status quo, and a policy-maker is now interested in counterfactual local
average treatment effects under intermediate subsidies.  
In particular, consider a policy that is expected to reduce treatment takeup by,
say, 10\% among experimental compliers. 
Information about conditional means does not bound the local average treatment
effect under the policy in the latent index selection model without further
structure or assumptions. 
Nevertheless, the model imposes such bounds because it identifies the marginal
distributions of potential outcomes among compliers.   
What matters then for the policy-relevant treatment effect is which 10\%
of compliers ``drop out'' of treatment relative to the experiment subsidy. 
In one extreme, the dropouts are those in the highest decile of the complier
treated outcome distribution and in the lowest decile of the untreated outcome
distribution; in the other extreme, the reverse is true.
This pair of extreme scenarios yield the sharp bounds on the effect of the
policy because the model does not impose any bounds on the joint distribution of
complier outcomes beyond consistency with the identified marginals. 

This basic intuition of ``trimming'' identified outcome distributions is related
to Lee~\cite{lee}.  
In a model of sample selection where a randomized treatment shifts sample
selection monotonically, he derives sharp bounds for the average treatment
effect among the always-sampled by thus bounding their average treated outcome. 
One significant difference in the present model with imperfect compliance is
that the trimming intuition is ``two-sided'' because neither average potential
outcome need be point-identified among a given subgroup.

The intuition is also generalized to analytically obtain sharp bounds on a
class of weighted average treatment effects --- technically, linear functionals
of the $\mte$ function --- via the theory of stochastic orders.%
\footnote{
    Given Vytlacil's~\cite{vytlacil} equivalence between monotonicity and a
    latent index selection model, this approach could also be useful for
    bounding a similar class of treatment effects in the sample selection
    framework of Lee~\cite{lee} when the selection equation is more explicitly
    modeled. 
}
More specifically, the sharp set of possible marginal treatment effect functions
is characterized in terms of a mean consistency and a second-order stochastic
dominance condition relative to the distribution of countermonotonic treatment
effects among compliers (\autoref{thm:sharp-set}). 
The importance of the countermonotonic effect (among compliers) is that it pairs
the highest treated outcomes with the lowest untreated outcomes and vice versa,
thus yielding the extremal treatment effect distribution consistent with the
identified marginals from which all candidate $\mte$ functions can be generated. 
Sharp lower and upper bounds on weighted average treatment effects are obtained
by rearranging an extremal marginal treatment effect function to be co- or
counter-monotonic with the weighting function
(\autoref{thm:sharp-te}).
The first result relies on properties of the convex stochastic order, and the
second also on properties of the supermodular stochastic order.%
\footnote{
    For a textbook reference on stochastic orders, see Shaked and
    Shanthikumar~\cite{shakedshanthikumar}. 
}

These results also relate and contribute to a literature on the identification
of the treatment effect distribution and its functionals in the case of perfect
compliance [Heckman et al.~\cite{hsc1997}, Fan and Park~\cite{fanpark2010},
Firpo and Ridder~\cite{firporidder2019}], because in this case the quantile
function of treatment effects constitutes a possible $\mte$ function.
I elaborate on this connection in \autoref{sec:related} after a presentation of
the main results. 
Furthermore, these results naturally suggest a complete graphical representation
in terms of uniformly sharp bounds (\autoref{thm:sharp-avg}), which relates the
quality of counterfactual inference to a well-known measure of statistical
dispersion (the Gini coefficient) and yields a specification test for further
parametric assumptions.%
\footnote{
    I focus on the recent linear parameterization of Brinch et
    al.~\cite{brinch}, which is an instance of a broader control function
    approach [Heckman and Robb~\cite{heckmanrobb}, Bj\"orklund and
    Moffitt~\cite{bjorklundmoffitt}; see also Kline and
    Walters~\cite{klinewalters}]. 
}

A central feature of the basic latent index selection model is its
lack of assumptions on the relationship between selection and potential
outcomes.  
Unsurprisingly, this can translate into weak inference over counterfactual
parameters,
especially as they become increasingly different from the ones
point-identified by the observed data.    
For example, reconsider the problem of bounding the effects of intermediate
subsidies.  
The data-generating processes that attain the non-parametric bounds exhibit two
extreme properties, which may be implausible in many applications.  
First, potential outcomes among compliers are perfectly countermonotonic:
those with the highest realized outcomes if treated would have obtained the
lowest outcomes if not treated, and vice versa. 
Second, compliers (negatively) select perfectly into treatment based on the resulting
extremal treatment effects.  
Additionally, the basic model does not provide information on counterfactual
treatment effects beyond compliers because one of the underlying potential
outcomes is always unobserved. 
Economic theory suggests a number of auxiliary assumptions a researcher may
consider in order to sharpen the non-parametric bounds.  
In their recent contribution, Mogstad et al.~\cite{mogstad} incorporate a
variety of conditional mean assumptions into their computational framework in
order to tighten partial identification. 

Another contribution of this paper is that the basic $\mte$ characterizations
can be used to incorporate and exhaust the content of additional distribution
assumptions, as evidenced by two simple extensions.%
\footnote{
    In a previous draft [Marx~\cite{marx2019}], I also propose a method for
    sharply incorporating conditional mean assumptions into the fully
    independent latent index selection model, using the representation of
    \autoref{thm:sharp-avg}.  
}
In the first extension, I aggregate over observed covariates and provide a
sharp policy-relevant interpretation of the aggregated bounds. 
Covariate information improves identification using marginal distributions 
by restricting the set of feasible joint distributions of potential outcomes. 
In the second extension, I sharply incorporate the rank similarity assumption of
Chernozhukov and Hansen~\cite{chernozhukovhansen}.   
Incorporating this assumption into the latent index framework confers several
advantages. 
First, the latent index framework defines a variety of policy-relevant treatment
effects, which are typically not point-identified under rank similarity alone,
even when the unconditional potential outcome distributions are
point-identified.    
Second, the added structure of the framework permits an alternative
identification strategy --- intuitively, extrapolation from compliers --- that
relaxes the need for continuity of outcomes in the existing method.  
This is important for the subsequent empirical application, in which some
outcomes are discrete and all outcomes have a mass point at zero.  
Third, the added empirical content of the rank similarity assumption is again
summarized compactly and graphically in terms of uniformly sharp bounds [as in
\autoref{thm:sharp-avg}], even though conditional marginal distributions
may no longer be fully identified.  
Thus the framework provides simple ways to compare or combine further
assumptions with rank similarity, as well as providing an alternative test of
(unconditional) rank similarity, as in Dong and Shen~\cite{dongshen} and
Frandsen and Lefgren~\cite{frandsenlefgren2018}. 

Empirically, I apply the methods using publicly available data from the Oregon
Health Insurance Experiment (OHIE) [Finkelstein~\cite{finkelstein-data}].
The OHIE, which extended Medicaid to a randomly selected group of uninsured,
low-income adults in Oregon, is unique in the experimental evidence it provides
on the impacts of insurance coverage on health outcomes. 
Yet many questions of policy interest --- for example, the effects of broader
expansions, encapsulated in proposals of ``Medicare for All'' --- require
extrapolation of point-identified treatment effects to different subpopulations.     
Specifically, I revisit the question of how insurance coverage affects emergency
room (ER) utilization. 
Previous analyses using the OHIE data find a positive effect of health insurance
on ER visits across a range of complier subpopulations [Taubman~\cite{taubman}],
yet predict negative effects across various measures of ER utilization for
expansions of Medicaid to the never-treated [Kowalski~\cite{kowalski2016}]; 
these predictions are based on the linear extrapolation method of Brinch et
al.~\cite{brinch}.  
My main empirical contribution is to show that an extrapolation based on the
distribution assumption of rank similarity 
predicts positive (in one case, nonnegative) average treatment effects among the
never-treated, which for some outcomes are even larger than the average
treatment effect among compliers. 
Thus distribution information and assumptions may provide substantively
different predictions than the ones obtained from mean information and
assumptions alone. 

The paper proceeds as follows. 
\autoref{sec:framework} reviews the latent index selection model and the
identification problem.
\autoref{sec:results} provides the main results and relates them in more detail
to the previous literature.
\autoref{sec:extensions} extends the results to auxiliary information and
assumptions.  
\autoref{sec:ohie} provides an empirical application using the publicly
available data from the Oregon Health Insurance Experiment.
\autoref{sec:conclusion} concludes. 

\section{Framework} 
\label{sec:framework} 

\subsection{Model}
In what follows, random variables are defined on a common probability space
$(\Omega, \F, P)$ with (cumulative) distribution function $F$. 
For a given random variable $X$, 
let $Q_X: [0,1] \to \mathds{R}$ denote its quantile function, $Q_X (q) = \inf \{
x: F_X (x) \geq q \}$, and let $I_X: [0,1] \to \mathds{R}$ denote its
\emph{integrated} quantile function (IQF), $ I_X (q) = \int_0^q \, Q_X (s) \,
ds$.  
A random variable $X$ is said to second-order stochastically dominate another
random variable $X'$, written $X \succeq_{SSD} X'$, if $I_X (q) \geq I_{X'} (q)$
for all $q \in [0,1]$.%
\footnote{
    This definition is equivalent to the perhaps better-known SSD
    characterization of Rothschild and Stiglitz~\cite{rothschildstiglitz70} in
    terms of a pointwise ordering of integrated \emph{distribution} functions,
    $\int_{-\infty}^x F_X (r) \, dr$.
    Equivalence follows from two facts: i) the integrated distribution and
    integrated quantile functions are convex conjugates of one another [Ogryczak
    and Rusczczy\'nski, \cite{or2002}] and ii) convex conjugation is
    order-reversing.
}
Two random variables $X$ and $X'$ defined on the same probability space
are comonotonic if $(X, X') \stackrel{d}{=} (Q_X (V), Q_{X'} (V))$ and
countermonotonic if $(X, X') \stackrel{d}{=} (Q_X (V), Q_{X'} (1-V))$, 
where $(X,X')$ denotes the random vector, $\stackrel{d}{=}$ denotes equality in
distribution, and $V \sim U[0,1]$.%
\footnote{
    Equivalently, the two random variables are comonotonic if:
    \[
        F_{(X,X')} (x,x') = \min \{ F_{X} (x), F_{X'} (x') \}
    \]
    and countermonotonic if: 
    \[
        F_{(X,X')} (x,x') = \max \{ F_X (x) + F_{X'} (x') - 1, 0 \}
    \]
    Note these are the upper and lower Fr\'echet-Hoeffding copula bounds,
    respectively.
}
In words, comonotonic random variables can be thought of as nondecreasing
functions of a single random variable $V$ and thus exhibit perfect positive
dependence; analogously, countermonotonic random variables exhibit perfect
negative dependence. 
The second-order stochastic dominance relation and monotonic random variables
will play an important role in subsequent characterizations. 

Proceeding with the variables of the model, let $D$ denote the binary
\emph{realized treatment decision} and let $Z$ denote an exogenous correlate of
the treatment decision, or \emph{instrument}.
For simplicity of notation and to crystallize the intuition of the partial
identification approach, I henceforth assume that the instrument $Z$ is also
binary.
For example, this occurs in policy evaluation settings with noncompliance, e.g.
\autoref{sec:ohie}.
In this case $Z$ is the treatment assignment, and I henceforth adopt this
terminology. 
Let $D_z$ denote the \emph{potential treatment decision} that would be
observed if the treatment assignment were exogenously set to $z$. 
Note that the realized treatment decision coincides with the potential decision
for the realized treatment assignment, $D = D_Z$.  
Let $\bar{p}_z = \E [ D | Z = z]$ denote the \emph{propensity score} (or choice
probability) as a function of treatment assignment. 

Let $Y$ denote the \emph{realized outcome}, and let $Y_{d,z}$ denote the
\emph{potential outcome} that would be observed if the treatment decision and
assignment were exogenously set to $(d,z)$. 
Henceforth I proceed under the exclusion restriction that the treatment
assignment affects the outcome only through its effect on the treatment
decision. 
Therefore suppress the assignment subscript $z$ and denote the potential
outcomes by $Y_d$. 
As previously, note that the realized outcome coincides with the potential
outcome for the realized treatment decision, $Y = Y_D$.  
Let $\Delta = Y_1 - Y_0$ denote the \emph{treatment effect}. 
The treatment effect
is the primary object of interest to the researcher.  

An identification challenge exists because the researcher only observes the
treatment assignment and the realized treatment decision and outcome.
\begin{assumption}[Data]
    \label{assn:data}
    The researcher observes the distribution of $(D,Y,Z)$. 
\end{assumption}

\noindent
In other words, for (almost) every individual, the researcher observes either
the treated outcome or the untreated outcome.   
Furthermore, the outcomes and the treatment effect may be correlated with the
treatment decision through unobservables $U$. 

The latent index selection model [Heckman and Vytlacil~\cite{hv1999,hv2005}]
non-parametrically identifies a variety of weighted average treatment effects
with a sufficiently rich source of variation. 
As shown by Vytlacil \cite{vytlacil}, its basic assumptions are equivalent to
those of the LATE model [Imbens and Angrist~\cite{imbensangrist}].  
However, the selection model's structural underpinnings make it particularly
useful for conducting counterfactual policy inference. 

\begin{assumption}[Model: Latent Index Selection with a Binary Instrument] { \quad }  
    \label{assn:lism}  
    \begin{enumerate} 
        \setcounter{enumi}{-1}
        \item \textbf{Binary instrument:} $P (Z = z ) > 0$ for $z \in \supp Z =
            \{0,1\}$.
        \item \textbf{Latent index rule:} Observed and counterfactual treatments
            are determined by an additively separable index rule. 
            For a standard uniform unobservable $U \sim U[0,1]$, 
            \begin{equation}
                \label{eq:selection}
                D_z = \mathds{1} [ U \leq \bp_z ] 
            \end{equation}
        \item \textbf{Independence:} $(U, Y_0, Y_1) \perp Z$. 
        \item \textbf{First stage:} $\bp_0 < \bp_1$. 
        \item \textbf{Expectations exist:} Potential outcomes $Y_0$ and $Y_1$ are integrable. 
    \end{enumerate}
\end{assumption}

\noindent
\autoref{assn:lism} constitutes the basic formulation of
Vytlacil~\cite{vytlacil} in the case of a binary instrument, with the
normalization that unobservables $U$ follow a standard uniform distribution.  
As discussed in Vytlacil~\cite{vytlacil}, this normalization is technically
innocuous given the other assumptions; at the same time, it underlies the
marginal treatment effect approach to identification introduced by Heckman and
Vytlacil~\cite{hv2005} and pursued in this paper. 
I now turn to this approach. 

\subsection{Target Parameters} 

Many causal questions of policy relevance have answers in terms of (weighted) averages of
the treatment effect $\Delta$ across some group of unobservables $\U \subseteq
[0,1]$. 
For a given unobservable $\U = \{ u \}$, Heckman and Vytlacil~\cite{hv2005}
define the marginal treatment effect: 
\begin{equation}
    \label{eq:mte}
    \mte (u) = \E [ \Delta | U = u].   
\end{equation}
The $\mte$ function has choice-theoretic foundations and is of interest in its
own right. 
For example, in the OHIE application of \autoref{sec:ohie}, $\mte(u)$ is the
expected change in spending (ER charges) from health insurance coverage for
unobservable type $U=u$.  
In the health insurance literature, this has been interpreted as the \emph{moral
hazard} of type $U=u$ [Einav et al.~\cite{einav}; Kowalski~\cite{kowalski2016}]. 
To the extent that the measured outcome proxies for the outcomes that matter to
the decision-maker, the $\mte$ may guide the treatment decision.  

The $\mte$ function also serves as a building block for other treatment effects.  
In particular, a wide array of policy-relevant average treatment effects can be
expressed as weighted averages, or linear functionals, of the $\mte$ function: 
\begin{equation}
    \label{eq:te}
    \te (w) = \int_0^1 \mte (u) \, w (u) \, du
\end{equation}
for some bounded weighting function $w : [0,1] \to \mathds{R}$ [Heckman
and Vytlacil~\cite{hv2005}]. 
Since unobservables are ordered according to their propensity for treatment via
\eqref{eq:selection}, a simple yet important family of groups $\U$ are the
intervals $[u,u']$. The corresponding set of treatment effects are the family of
\emph{local average treatment effects}: 
\begin{equation}
    \late (u,u') = \E [ \Delta | U \in [u,u']]
    \label{eq:late}
\end{equation}
which are obtained from \eqref{eq:te} by the weighting function $w_{u,u'} (s) =
\I (u \leq s \leq u') / (u'-u)$. 
As discussed by Heckman and Vytlacil~\cite{hv2005}, definition \eqref{eq:late}
generalizes the instrument-specific treatment effect introduced and identified
by Imbens and Angrist~\cite{imbensangrist}:  
\begin{equation}
    \late_{IA} 
    = \E [ \Delta | D_0 < D_1 ]
    = \late (\bp_0, \bp_1 ) 
    \label{eq:late-ia}
\end{equation}
to include effects that are not identified by the instrument at hand, but may
nevertheless be of policy interest. 
Perhaps the most widely studied is the \emph{average treatment effect} for the
population: 
\begin{equation}
    \ate 
    = \E [ \Delta ] 
    = \late (0,1) 
    \label{eq:ate}
\end{equation}
Beyond the $\late$ family, other treatment effects expressible as \eqref{eq:te}
include the average treatment effect on the treated, $\U = \{ U: D = 1 \}$.%
\footnote{
    The fact that $ D $ is measurable with respect to $U$ follows from the
    selection equation \eqref{eq:selection}. 
}
All aforementioned effects would be identified via \eqref{eq:te} if the $\mte$
function were identified.   

The $\mte$ function is not fully identified 
in the basic model with a discrete source of variation. 
Yet there is empirical content; for example, certain functionals of the $\mte$,
e.g. \eqref{eq:late-ia}, are identified [Imbens and Angrist~\cite{imbensangrist}]. 
Let $\M^*$ denote the set of candidate $\mte$ functions that are consistent with
the data (\autoref{assn:data}) and model (\autoref{assn:lism}). 
This set is \emph{sharp} with respect to these assumptions. 
In what follows I assume that the data-generating model is correctly specified,
so that the $\mte$ function exists and belongs to the set $\M^*$. 
Given the candidate set $\M^*$, a treatment effect $\te (w)$ of form
\eqref{eq:te} belongs to the set:
\begin{equation}
    \label{eq:te-problem}
    \left\{ 
        \int_0^1 m (u) w(u) \, du: m \in \M^*
    \right\}.
\end{equation}
Thus $\te (w)$ is (perhaps infinitely) bounded above and below by:  
\begin{equation}
    \label{eq:te-bounds}
    \inf_{m \in \M^*} \int_0^1 m(u) w(u) \, du 
    \, \leq \, \te (w) \, \leq \, 
    \sup_{m \in \M^*} \int_0^1 m(u) w(u) \, du 
\end{equation}
If $\M^*$ is a convex set, then the set of possible $\te (w)$ is an interval,
characterized by these upper and lower bounds. 
In this case, the bounds \eqref{eq:te-bounds} are \emph{sharp} for $\te (w)$
with respect to the model and data. 
Deriving the sharp bounds on $\mte$ and the class of $\te$ functionals in the
latent index model are the main theoretical contributions of this paper, which I
turn to after a preliminary definition and result. 

\section{Results} 
\label{sec:results}

\subsection{Preliminaries}
To state and prove the main results, I first define conditional random variables
based on a common partition of the population and then compile some related
identification results. 
Namely, let: 
\begin{equation}
    \label{eq:strata}
    G = \begin{cases} 
        a & \text{if $D_0 = D_1 = 1$} \\        
        c & \text{if $D_0 < D_1$} \\             
        n & \text{if $D_0 = D_1 = 0$} \\        
        d & \text{if $D_0 > D_1$}. 
    \end{cases} 
\end{equation}
denote a random variable that \emph{g}roups (or stratifies, in the language of
Frangakis and Rubin~\cite{frangakisrubin}) the population based on the
potential treatment decisions for each treatment assignment; 
the names $\{a,c,n,d\}$ stem from the \emph{a}lways-takers, \emph{c}ompliers,
\emph{n}ever-takers, and \emph{d}efiers nomenclature of Angrist, Imbens, and
Rubin~\cite{angristimbensrubin}.  
\autoref{assn:lism} rules out defiers, $\{G = d\} = \emptyset$, which are
henceforth omitted. 
For each other group $g$, let $(U_{g}, Y_{0,g}, Y_{1,g}, \Delta_g) \sim ((U,Y_0,
Y_1, \Delta) | G = g)$ denote a
random vector of eponymous random variables conditional on group membership $g$; 
these are well-defined for groups where $P (G = g) > 0$.  
The following preliminary lemma compiles known results about identification
within groups in terms of the conditional random variables.   

\begin{lemma}[Identification of Conditional Marginal Distributions]  
    \label{thm:group-lemma}
    The model and data identify all probabilities $P(G=g)$, and 
    the marginal distributions of conditional unobservables $U_g$ and 
    outcomes $Y_{d,g}$ for $d = 0, g \in \{c,n\}$ and $d = 1, g \in \{a,c\}$
    such that $P(G=g) > 0$.
    Conversely, any joint distribution of integrable
    $(\hat{U},\hat{Y}_0,\hat{Y}_1)$ consistent with 
    all identified marginals of $U_g$ and $Y_{d,g}$ 
    is consistent with the model and the data.
\end{lemma}

Identification of any group-conditional treatment effect $\Delta_g$ is notably
absent from \autoref{thm:group-lemma}. 
Instead, define $\Delta_g^-$ to be the group-conditional treatment effect if
outcomes $Y_{0,g}$ and $Y_{1,g}$ were 
countermonotonic: that is, higher instances of $Y_{1,g}$ were paired with lower
instances of $Y_{0,g}$. 
In contrast to the distribution of the \emph{true} group-conditional treatment
effect $\Delta_g$, the distribution of the extremal, countermonotonic treatment
effect $\Delta_g^-$ can be recovered from the marginal distributions of
$Y_{0,g}$ and $Y_{1,g}$, perhaps most simply through their quantile functions:
\begin{equation} \label{eq:Qneg} 
    \Delta_g^- \stackrel{d}{=} Q_{Y_{1,g}} (V) - Q_{Y_{0,g}} (1-V) 
    \quad \text{for $V \sim U[0,1]$} 
\end{equation}
The countermonotonic treatment effect plays an important role in the subsequent
partial identification results. 

\subsection{Sharp Set and Bounds} 
\label{sec:sharp}

This subsection provides the main partial identification results for the
marginal treatment effect ($\mte$) function and the class of weighted average
treatment effects.  
As is to be expected, meaningful partial identification is only possible in the
basic latent index model for the interval of compliers, where there is some
information about both treated and untreated outcomes.  
However, the extension in \autoref{sec:ext-dist} shows how these identification
results can nevertheless yield meaningful and sharp extrapolation beyond
compliers when additional assumptions are made.   

The main result characterizes the set $\M^*$ of candidate $\mte$ functions that
are consistent with the data and model.%
\footnote{
    Elaborating on the parenthetical condition, the data can only constrain
    feasible $\mte$ functions up to sets of measure zero, and occasionally (e.g.
    \autoref{sec:sum-bounds}) it will be helpful
    to think of $\M^*$ instead as a set of equivalence classes of integrable
    functions.
    In that case $\M^*$ is also a subset of the normed vector space of
    integrable functions $L^1 [0,1]$. 
}

\begin{theorem}[Sharp Set of the Marginal Treatment Effect Function] 
    \label{thm:sharp-set}
    Under \autoref{assn:data} and \autoref{assn:lism}, the sharp set $\M^*$ of
    the function $\mte(u) = \E [ \Delta | U = u]$, $u \in [0,1]$, consists of
    (equivalence classes of) integrable functions $m: [0,1] \to \mathds{R}$ that
    satisfy:
    \begin{align}
        \E [ m (U_c) ] &= \late_{IA} \label{eq:mean} \\[0.05in] 
        m (U_c) &\succeq_{SSD} \Delta_c^- \label{eq:ssd}
    \end{align}
    The sharp set $\M^*$ is convex. 
\end{theorem}

\noindent
\autoref{thm:sharp-set} splits the characterization of $\mte$ functions into two
conditions.
The first condition, as is well-known, requires consistency of the $\mte$
function with the identified $\late_{IA}$.
Adding the second condition requires further consistency of the $\mte$ function
with the distribution of countermonotonic treatment effects identified on the complier
interval. 

Intuitively, necessity of the condition follows in two parts.  
First, by definition of the $\mte$ as a function of conditional expectations of
treatment effects, the true group-conditional distribution of treatment effects
$\Delta_g$ is a mean preserving spread%
\footnote{In the sense of Rothschild and Stiglitz \cite{rothschildstiglitz70}.}
of the $\mte$ among that group: 
\begin{align}
    \E [ \mte (U_g) ] &= \E [ \Delta_g] \label{eq:mean-true} \\ 
    \mte (U_g) &\succeq_{SSD} \Delta_g \label{eq:ssd-true}. 
\end{align}
The first condition \eqref{eq:mean-true} implies \eqref{eq:mean}, but the true
group-conditional distribution of treatment effects $\Delta_g$ in
\eqref{eq:ssd-true} is typically unidentified --- even among compliers ---
because the model and data only identify (some) marginal distributions of
$Y_{d,g}$.   
However, each true group-conditional treatment effect is stochastically ordered
relative to the group-conditional countermonotonic treatment effect, $\Delta_g
\succeq_{SSD} \Delta_g^-$. 
Among compliers, where the countermonotonic treatment effect is identified,
combinining with \eqref{eq:mean-true} implies the necessity of \eqref{eq:ssd}. 

Conversely, the countermonotonic treatment effect $\Delta_c^-$ is always consistent
with the model and data, and any feasible $\mte$ function on the complier
interval can be constructed as if from this extremal distribution, even if that
distribution differs from the true distribution $\Delta_c$. 
Furthermore, there is no restriction (beyond integrability) on the $\mte$
function outside the complier interval because one of the potential outcomes is
unobserved. 

The next result identifies candidate $\mte$ functions that attain the sharp
bounds for the class of treatment effects $\te (w)$ whenever finite bounds
exist. 
These bounding solutions are characterized by two features.
First, they are extremal, in the sense that the distribution of values they take
among compliers is equal to the extremal distribution of feasible treatment
effects $\Delta_c^-$ identified by \autoref{thm:sharp-set}. 
Second, they are monotonic with respect to the weighting function.

\begin{theorem}[Sharp Bounds on $\te$ Functionals] 
    \label{thm:sharp-te}
    Consider a bounded weighting function $w: [0,1] \to \mathds{R}$. 
    Under \autoref{assn:data} and \autoref{assn:lism}, if the sharp bounds on
    $\te (w)$ are finite, then they are attained by a pair of solutions
    $\underline{m}_w, \overline{m}_w$ satisfying: 
    \begin{equation}
        \underline{m}_w (U_c), \overline{m}_w (U_c) \stackrel{d}{=} \Delta_c^- 
        \quad \text{and} \quad
    \begin{array}{l}
        w(U_c) \text{ and } \overline{m}_w (U_c) \text{ are comonotonic} \\[0.05in]
        w(U_c) \text{ and } \underline{m}_w (U_c) \text{ are countermonotonic} 
    \end{array} 
        \label{eq:functional}
    \end{equation}
\end{theorem}

\noindent
The basic intuition of \autoref{thm:sharp-te} is as follows. 
First, the integrand of $\te (w)$ defined in \eqref{eq:te} is a supermodular
function of $\mte$ and the weighting function $w$ (it is just the multiplication
of the two).  
Therefore, the value of $\te (w)$ is weakly increased (decreased) by
rearranging values of $\mte$ to match with weights in ascending (descending)
order.
Furthermore, the sharp set of $\mte$ in \autoref{thm:sharp-set} is closed under
such rearrangement operations within group $g$ because further information about
treatment propensity is unobserved.
This yields monotonicity in \eqref{eq:functional}. 
Second, the value of $\te (w)$ is increasing (decreasing) in mean-preserving
spreads of comonotonic (countermonotonic) $\mte$.   
Applying this intuition within the group of compliers --- where an extremal
distribution of treatment effects is identified --- yields the distribution of
the bounding solutions in \eqref{eq:functional}.  
Finally, finite bounds exist in the basic model only for weighting functions
that are essentially zero outside the interval of compliers. 
This is why the conditions on extremizing candidate functions in
\eqref{eq:functional} restrict their scope to compliers. 

An important special case of \autoref{thm:sharp-te} is the family of
counterfactual $\late$ parameters. 
Here one obtains the intuitive result that some parameters which are ``almost''
the same as $\late_{IA}$ are ``almost'' identified. 
In particular, such parameters are sharply bounded by the means of the extremal
distributions where the highest or lowest quantiles are trimmed off;
as explained in the introduction, this trimming intuition is reminiscent of
Lee~\cite{lee} in a related model with sample selection.
The bounds can be conveniently expressed using the integrated quantile
function.%
\footnote{
    In turn, the integrated quantile function of an integrable random variable
    $X$ can be obtained by cumulatively integrating over its quantile function,
    which is computationally straightforward. 
    If $X$ is continuous, then the function also has a representation in terms
    of a quantile-truncated expectation, $I_X (q) = q \E [ X | X \leq Q_X (q)]$;
    e.g.~see Jewitt~\cite{jewitt}.
}  

\begin{corollary}[Sharp Bounds on Counterfactual $\late$] 
    \label{thm:sharp-late}
    Under \autoref{assn:data} and \autoref{assn:lism}, the counterfactual
    $\late$ is bounded for any subinterval of compliers. 
    For $[u, u+\delta] \subseteq [\bp_0, \bp_1]$ and $\delta_c = \delta / P(G =
    c) = \delta / (\bp_1 - \bp_0)$, 
    \begin{equation}
        I_{\Delta_c^-} \left( \delta_c \right)
        \leq
        \delta_c \cdot \late (u, u+\delta)
        \leq 
        \late_{IA} - I_{\Delta_c^-} (1 - \delta_c)
        \label{eq:sharp-bounds-late}
    \end{equation}
\end{corollary}

\noindent
The result follows from \autoref{thm:sharp-te} given the $\late$ weighting
function $w_{u,u+\delta} (s) = \I (u \leq s \leq u+\delta) / \delta$. 
In particular, this indicator function places equal, positive weight on the
interval $[u, u+\delta] \subseteq [ \bp_0, \bp_1]$ and zero weight elsewhere. 
Therefore, by \autoref{thm:sharp-te}, the upper (lower) bounds are attained by
candidate $\mte$ functions that have the highest (lowest) values of the complier
countermonotonic treatment effect $\Delta_c^-$ arranged on the interval $[u,
u+\delta]$, similarly to the trimming intuition of Lee~\cite{lee}. 
Evaluating yields the bounds in \eqref{eq:sharp-bounds-late}. 

Since the value of the $\mte$ function at a point $u$ can be almost everywhere
recovered
as the limit of $\late (u, u+\delta)$ where $\delta \to 0$, it may appear that
\autoref{thm:sharp-late} also imposes finite pointwise bounds on the $\mte$
function at some points.  It does not. 
Moreover, it cannot because a given point is infinitesimal, and the model and
data do not impose any constraints on the $\mte$ function for sets of measure
zero.
However, the following subsection shows how the model imposes not just pointwise
but (among compliers) \emph{uniformly} sharp bounds on a closely related object,
namely the counterfactual average outcome as a function of the latent index
treatment threshold.   
In turn, these pointwise bounds provide a simple, graphical way to summarize the
sharp bounds derived so far. 

\subsection{Summarizing the Bounds} 
\label{sec:sum-bounds}

This subsection summarizes the previous bounds in terms of pointwise (a
fortiori, uniformly) sharp bounds for a real-valued function on $\mathds{R}$.  
This yields a simple and new graphical representation of the latent index
model's empirical content for the $\mte$ function and its linear functionals.
To that end, define the \emph{counterfactual average outcome} $\by (p)$ to be
the average outcome that would occur if the threshold in the latent index
selection equation \eqref{eq:selection} were exogenously set to $p \in [0,1]$: 
\[
    \by (p) = p \E [ Y_1 | U \leq p] + (1-p) \E [ Y_0 | U > p]. 
\]
The counterfactual average function also has a representation in terms of the
$\mte$ function: 
\begin{equation}
    \by (p) = \E [Y_0] + \int_0^p  \mte (u) \, du. 
    \label{eq:int-mte}
\end{equation}
This well-known representation underpins the Local Instrumental Variable (LIV)
procedure for estimation of the $\mte$ function [Heckman and
Vytlacil~\cite{hv1999, hv2000}].  

The following result provides pointwise sharp bounds on counterfactual averages
for the interval of compliers, where finite bounds exist.  

\begin{corollary}[Pointwise Sharp Bounds on Counterfactual Averages] 
    \label{thm:bounds-cao} 
    Under \autoref{assn:data} and \autoref{assn:lism}, the counterfactual
    average is bounded pointwise on the interval of compliers.   
    For $p \in [\bp_0, \bp_1]$, 
    \begin{equation}
        \label{eq:sharp-avg}
        \psi_l (p) 
        \, \leq \, 
        \by (p) 
        \, \leq \, 
        \psi_h (p)
    \end{equation}

    \vspace{-0.15in}
    \noindent
    where: 

    \vspace{-0.25in}
    \[
        \begin{array}{rcl} 
            \psi_l (p) 
            &=&
            \by (\bp_0) + (\bp_1 - \bp_0) \cdot I_{\Delta_c^-} 
            \left( \frac{p - \bp_0}{\bp_1 - \bp_0} \right) \\[0.1in]
            \psi_h (p)
            &=&
            \by (\bp_1) - (\bp_1 - \bp_0) \cdot I_{\Delta_c^-} 
            \left( \frac{\bp_1 - p}{\bp_1 - \bp_0} \right)
        \end{array} 
    \]
    Else, for $p \notin [\bp_0, \bp_1]$, 
    \[
        \psi_l (p) < \by (p) < \psi_h (p) 
    \]
    where $\psi_l (p) = -\infty$ and $\psi_h (p) = \infty$. 
\end{corollary}

\noindent 
This result follows from \autoref{thm:sharp-late} upon substituting
$\late(\bp_0,p) = \frac{\by (p) - \by(\bp_0)}{p - \bp_0}$, $\late_{IA} =
\frac{\by (\bp_1) - \by(\bp_0)}{\bp_1 - \bp_0}$ and simplifying. 
Furthermore, the propensity scores $\bp_z = \E [ D | Z = z]$ and the mean
outcomes $\by (\bp_z) = \E [ Y | Z = z]$ are identified, so that
\eqref{eq:sharp-avg} indeed bounds the counterfactual average among compliers. 

The bounds of \autoref{thm:bounds-cao} on the counterfactual average function
encode the sharp set and bound results of \autoref{sec:sharp} through operations
of function \emph{rearrangement}.%
\footnote{
    The notion and language of rearrangement alludes to the close connection
    between the stochastic \emph{convex order} (at the heart of
    \autoref{thm:sharp-set}) and the function \emph{majorization} order
    (implicit in \autoref{thm:sharp-avg}).    
    Namely, random variables are ranked in the convex order if and only if their
    quantile functions are ranked in the majorization order;
    Furthermore, the quantile function effectively serves as a rearrangement
    operator. 
    The fundamental majorization result is due to Hardy, Littlewood, and
    P\'olya~\cite{hardylittlewoodpolya}. 
    A textbook reference (with a focus on the discrete theory) is Marshall et
    al.~\cite{marshallolkinarnold}. 
}
Say that functions $m, \tilde{m}:[0,1] \to \mathds{R}$ are \emph{rearrangements}
of one another if the distribution of values taken by the functions are the
same, $\tilde{m} (U) \stackrel{d}{=} m (U)$. 
As a special case, define the \emph{increasing rearrangement among compliers} of
an integrable function $m$ by:   
\[
    m^{\uparrow_c} (u) = 
    \begin{cases} 
        Q_{m (U_c)} \left( \frac{u - \bp_0}{\bp_1-\bp_0} \right) 
        & \text{if $u \in [\bp_0, \bp_1]$} \\
        m (u) & \text{if $u \notin [\bp_0, \bp_1]$} 
    \end{cases} 
\]
The transformation ${}^{\uparrow_c}$ rearranges the function's values on the
complier interval in increasing order via the quantile function of $m (U_c)$. 
Let $\A^*$ denote the sharp set of candidate counterfactual average outcome
functions;  
let $a$ denote a generic element and $a'$ its derivative (where it exists). 
The following result restates the empirical content of \autoref{thm:sharp-set}
in terms of rearrangements and the counterfactual average function. 

\begin{proposition}[Sharp Set of Counterfactual Average Outcomes]
    \label{thm:sharp-avg}
    Under \autoref{assn:data} and \autoref{assn:lism}, the sharp set $\A^*$ of
    counterfactual average functions consists of absolutely continuous functions
    $a: [0,1] \to \mathds{R}$ that satisfy:
    \begin{equation}
        \psi_l (p) 
        \leq
        a (\bp_0) + \int_{\bp_0}^p (a')^{\uparrow_c} (u) \, du
        \leq
        \psi_h (p) 
        \label{eq:sharp-cao}
    \end{equation}
    for the pointwise bounds $\psi_l, \psi_h$ of \autoref{thm:bounds-cao}.
    The sharp set $\A^*$ is convex. 
\end{proposition}

One implication of \autoref{thm:sharp-avg} is that the bounds $\psi_l$ and
$\psi_h$ in \eqref{eq:sharp-avg} are \emph{uniformly} sharp on the complier
interval: there exists a data-generating process consistent with
\autoref{assn:data} and \autoref{assn:lism} for which the true counterfactual
average outcome coincides with the upper (lower) bounding function on the entire
complier interval $p \in [\bp_0, \bp_1]$. 
Namely, the lower (upper) bounds are uniformly attained by a data-generating
process with i) perfect countermonotonicity of complier potential outcomes,
captured by $\Delta_c^-$, and ii) perfect negative (positive) selection into
treatment among compliers, captured by comonotonicity (countermonotonicity) of
$\Delta_c^-$ and $U_c$.
Thus the bounding functions are themselves segments of feasible average outcome
functions.  
Yet \autoref{thm:sharp-avg} also shows that satisfying the uniform bounds is not
sufficient for an absolutely continuous function to be a candidate
counterfactual average function;
the bounds must additionally be satisfied by the increasing (in fact, any)
rearrangement among compliers.  

The information encoded in the uniform bounds is sufficient to recover the sharp
bounds on functionals $\te (w)$ of \autoref{thm:sharp-te}. 
In particular, the derivative of the uniform lower (or upper) bound recovers the
distribution of the countermonotonic treatment effect $\Delta_c^-$ via its
quantile function: 
\[
    (\psi_l)' (p) = Q_{\Delta_c^-} \left( 
        \frac{p - \bp_0}{\bp_1 - \bp_0}
    \right)
    \quad \text{for almost every $p \in [\bp_0, \bp_1]$.} 
\]
Thus $(\psi_l)' (U_c) \stackrel{d}{=} \Delta_c^-$, and so the sharp bounds for
$\te (w)$ among compliers can be obtained by rearranging the extremal $\mte$
function $(\psi_l)' (u)$ to be co- and counter-monotonic with the weighting
function $w(u)$. 
Stated in reverse, the increasing rearrangement (among compliers) of candidate
$\mte$ functions that attain the sharp bounds coincides (among compliers $u \in
[\bp_0, \bp_1]$) with the segment of the extremal $\mte$ function $(\psi_l)'
(u)$. 
Analogous logic holds for decreasing rearrangements and the upper uniform bound. 
Thus the uniform bounds on the counterfactual average summarize the empirical
content of \autoref{assn:data} and \autoref{assn:lism} on treatment effects.

An appeal is that these bounds can be depicted graphically, as illustrated
empirically in \autoref{sec:ohie}. 
This graphical representation helps to visualize, summarize, and compare the
extrapolation power of the model and its special cases; additionally, as will be
discussed in remaining (sub)sections, it helps to assess the consistency of
additional assumptions.  

The area between the uniform bounds provides an intuitive  
and theoretically grounded summary statistic of the extrapolation power
conferred by the model and data. 
Namely, define the \emph{area of uncertainty} $R (p,p'): [0,1]^2 \to
\mathds{R}_+ \cup \{ \infty \}$ as the total area between the uniform bounds on an interval
$[p, p']$:
\[
    R (p, p') = \int_p^{p'} [ \psi_h (u) - \psi_l (u) ] \, du
\]
and define the \emph{average area of uncertainty} for $p' > p$ as $\bar{R}
(p,p') = R(p,p') / (p' - p)$.
For any integrable random variable $X$, define its Gini mean difference
$\Gamma_X$ and, if $\E [ X ] \neq 0$, its Gini index $\gamma_X$ by:%
\footnote{
    The IQF representation of the Gini index can be found in Muliere and
    Scarsini~\cite{mulierescarsini}. 
    Note that if $X$ takes negative values, the Gini index may be greater than
    one. 
}
\begin{equation}
    \label{eq:gini} 
    \Gamma_X = 2 \int_0^1 [ q \E [ X ] - I_X (q) ] \, dq, 
    \quad \quad 
    \gamma_X = \frac{ \Gamma_X }{ \E [X] } 
\end{equation}
The next result ties the (average) area of uncertainty in the basic model to the
normalized and unnormalized Gini coefficients. 

\begin{proposition}[Area of Uncertainty] 
    \label{thm:gini}
    Under \autoref{assn:data} and \autoref{assn:lism}, the average area of
    uncertainty among compliers is finite and given by:  
    \[
        \bar{R} ( \bp_0, \bp_1 ) = (\bp_1 - \bp_0) \cdot \Gamma_{\Delta_c^-}
    \]
    If additionally $\late_{IA} \neq 0$, then the average area of uncertainty
    can also be expressed as: 
    \[
        \bar{R} ( \bp_0, \bp_1) = (\bp_1 - \bp_0) \cdot \late_{IA} \cdot
        \gamma_{\Delta_c^-} 
    \]
\end{proposition}

\noindent
In the case of a nonzero treatment effect, the average area of uncertainty among
compliers is a product of three features of the data: the first stage $\bp_1 -
\bp_0$, the complier treatment effect $\late_{IA}$, and the amount of
mean-normalized dispersion around the countermonotonic treatment effect, as
captured by the Gini index $\gamma_{\Delta_c^-}$.  

Despite the apparent simplicity of this expression, care is required in its
interpretation. 
Consider the size of the first stage. 
In addition to its direct effect on the average area of uncertainty, the size of
the first stage also has an indirect effect of affecting dispersion by reducing
uncertainty about the relationship between unobservables and potential outcomes. 
Thus, interpolation improves as the size of the first stage decreases (ignoring
any statistical complications), but at the expense of being confined to a
smaller interval. 
In contrast, even though $\late_{IA}$ appears in the second expression, it does
not have an effect on the average area of uncertainty per se.  
For example, 
adding a constant to the distribution of treatment effects would change
$\late_{IA}$ without any effect on the average area of uncertainty. 
Intuitively, this is because the term also enters in the denominator of
the Gini index.
Finally, the main determinant of the average area of uncertainty is the amount
of dispersion in the distribution of the countermonotonic treatment effect.  
This can be as low as zero if potential outcomes are constant, or arbitrarily
large for sufficiently dispersed distributions.

The (average) area of uncertainty is infinite among non-compliers in the basic
model, which intuitively reflects the fact that the basic model provides no
ability to reason counterfactually about treatment effects beyond compliers. 
Nevertheless, the average area of uncertainty remains well-defined as
assumptions are added to the model, and thus also provides a useful family of
summary measures for studying and comparing models within the latent index
selection framework. 

\subsection{Relation to the Literature}
\label{sec:related} 

\subsubsection{Bounds in a Mean-Independent Latent Index Selection Model} 

In a recent but already influential contribution, Mogstad et al.~\cite{mogstad}
introduce a computational framework for counterfactual inference in a
mean-independent latent index selection model. 
Their approach is based on i) 
a set of IV-like estimands that summarize the observed data, and ii) a set of
marginal outcome functions that can encode various kinds of additional
assumptions a researcher might wish to impose. 
Remarkably, even with the flexible nature of their framework, they show that
their bounds can exhaust the empirical content of means.
Thus their framework recovers the mean consistency condition \eqref{eq:mean} in
\autoref{thm:sharp-set}.

Yet mean consistency per se provides no counterfactual content in the latent
index selection model --- in other words, it places no restrictions on treatment
effects beyond $\late_{IA}$.   
In contrast, the distribution consistency condition \eqref{eq:ssd} of
\autoref{thm:sharp-set} shows that the (fully independent) latent index
selection model has some counterfactual content.    
It imposes bounds on treatment effects among subgroups of compliers; that is, it
permits partial interpolation.%
\footnote{
    An application suggested by Kowalski~\cite{kowalski2016} in the context
    of the Oregon Health Insurance Experiment (studied in \autoref{sec:ohie}) is
    that a policy-maker might consider providing lottery winners the opportunity
    to receive discounted rather than free insurance coverage.
    Then interpolated treatment effects from the experiment would be
    policy-relevant.  
}
Furthermore, full independence of the instrument is a common 
empirical assumption, which is frequently motivated by (quasi)random
assignment.    
Perhaps for that reason, despite being logically stronger, full independence is
also a common assumption in the theoretical literature.%
\footnote{
    E.g. Imbens and Angrist~\cite{imbensangrist} and Vytlacil~\cite{vytlacil}.
    Heckman and Vytlacil~\cite{hv1999, hv2005} assume full \emph{pairwise}
    independence $(Y_d, U) \perp Z$ for $d = 0,1$.
}

Another relative contribution of the present paper is that 
the sharp bounds are derived explicitly, rather than being characterized
implicitly as the optimal values to a pair of infinite-dimensional convex
programs akin to \eqref{eq:te-problem}. 
In fact, despite their infinite-dimensional nature, the solutions underlying the
sharp bounds can be described relatively simply.  
Such explicit sharp bounds help
the researcher reason about where they might be willing to entertain more (or
fewer) assumptions. 
For example, a particularly stark feature of the sharp bounds in the basic model
is the possibility of perfect countermonotonicity in the treatment effect among
compliers. 
In \autoref{sec:extensions} I show how the results presented so far can be
extended to accommodate a common distributional assumption at the other extreme,
namely rank similarity or invariance.
Thus the results of this paper can also be used as a basis for incorporating and
comparing additional assumptions.%

Finally, the sharp bounds can be used to test the validity of additional
assumptions, such as parameterizations of the $\mte$ function.  
A popular and natural candidate is the linear marginal outcome assumption of
Brinch et al.~\cite{brinch}, which point identifies the $\mte$ function with just
a binary instrument. 
In other words, linearity reduces the sharp set $\M^*$ to a singleton, and thus
achieves point identification of all weighted average treatment effects $\te
(w)$. 
The candidate $\mte$ function can be integrated to obtain a counterfactual
average function, for which the uniformly sharp bounds of
\autoref{thm:bounds-cao} provide a specification test in the latent index
selection model. 
In fact, consistency with the uniform bounds is also \emph{sufficient} in the
case of a linear $\mte$ candidate function.  
This follows because such a function is monotone, and therefore equal to either
its increasing or decreasing rearrangement among compliers. 
These observations are collected in the following result. 

\begin{corollary}
    \label{thm:linear-specification}
    A linear $\mte$ function is consistent with \autoref{assn:data} and
    \autoref{assn:lism} if and only if the corresponding counterfactual average
    function obtained via \eqref{eq:int-mte} 
    satisfies the uniformly sharp bounds of \autoref{thm:bounds-cao}.  
\end{corollary}

\noindent
As will be shown in \autoref{sec:extensions}, similar results can be obtained
for the \emph{joint} consistency of parameterizations with additional
assumptions, for example on the joint distribution of potential outcomes.   

\subsubsection{Bounds on Treatment Effects with Perfect Compliance}

The present work also relates to a literature beginning with Heckman et
al.~\cite{hsc1997} that studies the distribution of the treatment effect in
models of perfect compliance. 
Most relatedly, Fan and Park~\cite{fanpark2010} and Firpo and
Ridder~\cite{firporidder2019} study bounds on the distribution of the treatment
effect and functionals thereof.  
Fan and Park~\cite{fanpark2010} show that the treatment effect satisfies a
second-order stochastic dominance conditions relative to the comonotonic and
countermonotonic treatment effect: 
\begin{equation}
    \Delta^- \preceq_{SSD} \Delta \preceq_{SSD} \Delta^+ 
    \label{eq:te-ssd}
\end{equation}
In contrast to the $\mte$ function, however, the second-order stochastic
relation \eqref{eq:te-ssd} does not characterize the set of feasible treatment
effects. 
As shown by counter-example in \autoref{apx:te-example}, the same remains true
even when the second-order bounds are combined with the first-order
Makarov~\cite{makarov} bounds, also introduced in the treatment effect framework
by Fan and Park~\cite{fanpark2010}. 
Thus, the known bounds on the distribution of treatment effects are \emph{not
sharp} in the full sense, and characterizing the sharp set of the distribution
of treatment effects remains an open problem. 

It is interesting to contrast this conclusion with the results of the present
paper, which include a derivation of the sharp set on marginal treatment effects
in the latent index selection model. 
To simplify the comparison, I consider the special case of perfect compliance $D
= Z$ and drop the redundant complier subscript.  
In that case, the quantile function of the treatment effect $Q_\Delta (\cdot)$
is a feasible $\mte$ function, which corresponds to perfect (hypothetical)
negative selection on the treatment effect.  
Thus the necessary and sufficient conditions for a feasible $\mte$ function are
necessary conditions for a feasible quantile function. 
The conditions are not sufficient.

Rather, the $\mte$ function is akin to a \emph{fractional} quantile function,
and the problem of deriving sharp bounds on functionals of the $\mte$ is a
relaxation of the problem of deriving sharp bounds on functionals of the
quantile function.%
\footnote{
    The $\mte$ relaxation of the quantile problem bears a noteworthy resemblance
    to Kantorovich's linear programming relaxation of Monge's optimal transport
    problem.  
    For more detail in the context of economics, see Galichon~\cite{galichon}.  
}
Thus, if the sharp bounds on an $\mte$ functional 
are attained by feasible quantile functions, then the bounds are also sharp over
the set of feasible quantile functions.   
By \autoref{thm:sharp-set}, the relaxed problem has relatively useful features,
such as an analytical characterization and convexity of its feasible set. 
This suggests a potentially fruitful method for deriving bounds on functionals
of the quantile function, even without a characterization of the quantile
function sharp set. 
Even when the bounds are not attained by quantile functions, solutions to the
relaxed $\mte$ problem may be insightful, as illustrated by a simple example in
\autoref{apx:te-example}.

\section{Extensions} 
\label{sec:extensions}

\subsection{Covariates} 
\label{sec:covars}

I now discuss and extend the previous results in the case where
\autoref{assn:lism} holds conditional on additional observed covariates, denoted
by $X$. 
Covariate information is useful for at least two reasons. 
The first reason is that it permits subgroup analysis and comparison. 
Specifically,
the latent index rule \eqref{eq:selection} can be generalized to
accommodate covariates: 
\begin{equation}
    D_z = \mathds{1} [ U \leq \bp_{z,X} ],   
    \label{eq:selection-x}
\end{equation}
where $U \sim U [0,1]$, $U \perp X$, and $\bp_{z,x} = \E [ D | Z=z, X=x]$.
Then analysis in terms of a covariate-conditional marginal treatment effect
function $\mte (u,x) = \E [ \Delta | U = u, X=x]$ proceeds as previously.  
The second reason is that aggregation over covariates can be used to improve
unconditional inference.  
This will be the focus of the remaining discussion. 

In order to endow the improved bounds with policy relevance, consider a family
of alternative policy interventions $\W \subseteq \mathds{R}$ satisfying the
following assumptions. 

\begin{assumption}[Alternative Policy Interventions] {\quad} 
    \label{assn:policies}
    \begin{enumerate}
        \item \textbf{Policy Invariance:} $(Y_{0,w}, Y_{1,w}, U_w, X_w) = (Y_0,
            Y_1, U, X)$ for all $w \in \W$.
        \item \textbf{Policy Monotonicity:} 
            Potential treatment across policies is determined by: 
            \begin{equation}
                \tilde{D}_w = \mathds{1} [ U \leq \nu (w,X) ],  
                \label{eq:selection-pr}
            \end{equation}
            where $\nu (\cdot, x)$ is a nondecreasing, right-continuous function
            of $w$ for every $x$.   
        \item \textbf{Instrument Inclusion:} For each $z \in \{0,1\}$, there
            exists a policy $\bw_z \in \W$ such that: 
            \begin{equation} 
                D_z = \tilde{D}_{\bar{w}_z}. 
            \end{equation}   

        \item \textbf{Continuity:} The policy cutoff $W$ defined by:  
            \begin{equation}
                \label{eq:w} 
                W = \inf \{ w : U \leq \nu (w,X) \},  
            \end{equation} 
            is a continuous random variable. 
    \end{enumerate}
\end{assumption}

\noindent
Following a line of research beginning with Heckman and
Vytlacil~\cite{hv2001, hv2005}, policy invariance assumes that the
alternative policies affect only the propensity for treatment, and not the
potential outcomes, unobservables, or covariates.%
\footnote{
    Policy invariance in \autoref{assn:policies} is the same as in Carneiro et
    al.~\cite{carneiro2010, carneiro2011}.  
    It can be weakened to an assumption on the \emph{distribution} of these
    random variables across policies $w$; for more detail, see Heckman and
    Vytlacil~\cite{hv2005}.  
}
Policy monotonicity and instrument inclusion impose structure on how alternative
policies affect treatment relative to one another and relative to the observed
experiment; 
in this respect, the substantively novel structure relative to the existing
literature is the monotonicity across covariates.%
\footnote{
    \label{fn:policies}
    Conditional on covariates, policy monotonicity and instrument inclusion
    recover the essence of the 
    \emph{stochastic} policy definition of the aforementioned papers [Heckman
    and Vytlacil~\cite{hv2001,hv2005}, Carneiro et al.~\cite{carneiro2010,
    carneiro2011}], in which each policy $*$ is associated with a stochastic
    threshold $P^*$ such that $D^* = \mathds{1} [ U \leq P^* ]$. 
    The reason is that the stochastic policies are assumed to be exogenous,
    $(Y_0, Y_1, U) \perp P^*$, and so the expected outcome $\E [ Y^* ]$ 
    is equivalent to a mixture over uniform policy cutoffs in the population: 
    \[
        \E [ Y^* ] = \E_{P^*} [ P^* \E [ Y_1 | U \leq P^* ] + (1-P^*) \E
        [Y_0 | U > P^* ]]
    \]
    The conceptual distinction between the approaches is whether a policy is
    associated with a 
    shift of cutoffs relative to the observed instrument $Z$, or with a (mixture
    of) uniform cutoff(s) in the population relative to the
    \emph{counterfactual} instrument realization $z$; the exogeneity of the
    instrument and policy shifts under consideration makes the two
    computationally equivalent. 
}
However, covariate monotonicity appears reasonable for the typical policies that
also satisfy policy invariance, such as prices or subsidies for treatment takeup
in the absence of general equilibrium effects.%
\footnote{
    The direction of monotonicity across covariates can be inferred from the
    data. 
    In the case where this direction varies significantly with covariates, a
    similar analysis could be undertaken by
    splitting the space of covariate values into subsets of positive and
    negative response to the policies.
    See, e.g., Semenova~\cite{semenova2021} for an approach to flexibly
    accommodating such covariate-conditional monotonicity in the related sample
    selection framework of Lee \cite{lee}.
}
Finally, the policy cutoff $W$ satisfies $\tilde{D}_w = \mathds{1} [ W \leq w]$
for all $w \in \W$ by definition; continuity of the cutoff is a technically
convenient assumption that allows the cutoff to be redefined in terms of its
unconditional quantile $W := F_W (W)$,
so that $W \sim U[0,1]$ and $\bw_z = \E [ \bp_{z,X}]$ for $z \in \{0,1\}$.  
This normalization is adopted henceforth. 

Next, define the \emph{marginal policy relevant treatment effect} function:
\begin{equation}
    \mprte (w) = \E [ \Delta | W = w].  
\end{equation}
While this definition differs somewhat from the preceding literature,%
\footnote{
    Namely, Carneiro et al.~\cite{carneiro2010, carneiro2011} consider a
    sequence of stochastic exogenous policies $\{ P_w^* \}_{w
    \in \W}$ and define $ \mprte (w) = \lim_{w \to 0} \, ( \E [ Y_w^*] - \E
    [Y_0^*]) /  (\E [ D_w^*] - \E [ D_0^*])$, where $P_0^* = \bp_Z$.
    The difference in definitions is consistent with the distinction discussed
    in \autoref*{fn:policies}.
}
it yields a close parallel between the $\mte$ and $\mprte$ functions.
Of course, the connection between marginal treatment effects and policy
relevance is a primary motivation for the study of the $\mte$ function and its
functionals in the first place.     
Yet, the set of feasible $\mprte$ functions is additionally constrained relative
to the feasible characterization of \autoref{thm:sharp-set} because the extremal
treatment effects underlying the bounds can be ``at most'' perfectly
countermonotonic \emph{conditional on covariates}. 

Elaborating, let $ \Delta_{g,x}^-$ denote a random variable distributed
identically to the treatment effect restricted to and perfectly
countermonotonic within the stratum $(g,x)$.%
\footnote{
    As in \eqref{eq:Qneg}, such a random variable is most easily defined in
    terms of its quantile function. Letting $Y_{d,g,x} \sim (Y_d | G=g, X=x)$,
    \[
        \Delta_{g,x}^- \stackrel{d}{=} 
        Q_{Y_{1,g,x}} (V) - Q_{Y_{0,g,x}} (1-V) 
        \quad \text{where $V \sim U[0,1]$.} 
    \]

    \vspace{-0.15in}
}
Then define $\Delta_{g,\bx}^-$, the average group-conditional countermonotonic
treatment effect across $X$, by the distribution function: 
\begin{equation}
    F_{\Delta_{g,\bx}^-} (\delta) = \E [ F_{\Delta_{g,X}^-} (\delta) | G=g]  
    \label{eq:agg-x}
\end{equation}
Let $W_g$ denote the random variable of policy cutoffs constrained to $G=g$.
It follows analogously to the reasoning of \autoref{thm:sharp-set} that the set
$\R$ of feasible $\mprte$ functions is constrained to the set of integrable
functions $r: [0,1] \to \mathds{R}$ satisfying the stochastic relations: 
\[
    \E [ r (W_c) ] = \late_{IA} 
    \quad \text{and} \quad r (W_c) \succeq_{SSD} \Delta_{c,\bx}^- 
\]
Furthermore, since the assumptions on $\nu$ in \eqref{eq:selection-pr}
are agnostic about relative selection across covariates, this conditional
countermonotonicity is the only additional restriction imposed.   
Of course, further parametric restrictions on $\nu$ would tighten bounds.%
\footnote{
    For example, one might assume additive or proportional shifts in treatment
    takeup for policies across covariates. 
    This is similar in spirit to the examples studied in Carneiro et
    al.~\cite{carneiro2010, carneiro2011}, except that the restrictions are
    applied across (rather than conditional on) covariates. 
}
Similarly, as explored in the next extension, bounds can be tightened through
further assumptions on the joint distribution of potential outcomes. 

\subsection{Rank Similarity} 
\label{sec:ext-dist}

The next extension illustrates how the preceding framework and results can
be used to (sharply) accommodate additional assumptions on the joint
distribution of outcomes. 
Such assumptions address a stark feature underlying the sharp bounds in the
basic model: among compliers where the distribution of treated and untreated
outcomes is observed, treatment effects could be perfectly countermonotonic,
and among non-compliers, treatment effects are not restricted at all.  
Equally starkly, suppose that the joint distribution of potential
outcomes --- and thus the distribution of treatment effects --- were known.  
Such knowledge could significantly improve the quality of inference over
counterfactual weighted treatment effects $\te (w)$. 
Yet this still might not point identify all such effects because of the
remaining uncertainty about the relationship between treatment effects and
the propensity for selection into treatment. 

Rank similarity or invariance is a
strong but often reasonable assumption on the joint distribution of potential
outcomes that clearly illustrates these improvements and limitations.
The identifying power of both assumptions is studied previously by Chernozhukov
and Hansen~\cite{chernozhukovhansen}, who focus on identification of the
population potential outcome distributions in a generalized setting with
possibly non-binary treatments and a non-separable selection rule. 
Those weaker assumptions suffice for identifying population quantiles, but not
for identifying (let alone defining) policy-relevant parameters such as those 
derived from the $\mte$ function. 
This paper's contribution is to sharply incorporate the rank similarity and
invariance assumptions into the latent index selection framework, which allows
i) partial identification of weighted treatment effects $\te (w)$ beyond the
average treatment effect, and ii) the relaxation of a previous continuity
assumption that is violated in the empirical application of \autoref{sec:ohie}.%
\footnote{
    In other related work, Vuong and Xu~\cite{vuongxu} identify individual
    treatment effects (and thus also the average treatment effect) under rank
    invariance and the monotonicity assumption of Imbens and
    Angrist~\cite{imbensangrist}. 
    The key to their main result on point identification is the existence of a
    deterministic mapping between potential outcomes, which relies on continuity
    of potential outcome distributions.  
    The existence of mass points explains the difference in our results,
    including why the $\ate$ in my application (\autoref{sec:ohie}) is not
    point-identified.  
}

\begin{assumption}[Rank Similarity or Invariance] 
    \label{assn:rank}
    { \quad } 
    \begin{enumerate}
        \item[a.]  
            Potential outcomes satisfy rank similarity if for $d = 0,1$
            there exist rank variables $ V_d $ that satisfy $Y_d = Q_{Y_d}
            (V_d)$ and that are identically distributed conditional on $U$.%
            \footnote{
                As discussed by Chernozhukov and
                Hansen~\cite{chernozhukovhansen}, the quantile or Skorokhod
                representation is without empirical loss of generality because
                i) by \autoref{assn:data} the researcher only observes the
                distribution of data, and ii) for any distribution function such
                a random variable and representation exist; 
                see e.g.~p.~34 of Williams~\cite{williams}.
            }
        \item[b.] Potential outcomes satisfy rank invariance if the rank
            variables are equal: $V_0 = V_1$. 
    \end{enumerate}
\end{assumption}

Rank similarity is theoretically weaker than rank invariance because it allows
random slippage in ranks conditional on the selection unobservable $U$. 
However, the empirical content of the two assumptions is identical because there
always exist extremal data-generating processes under rank similarity that are
rank invariant.  
Specifically, define (on the original probability space) an ``as if'' comonotonic
treatment effect by:%
\footnote{
    Thus defined, the comonotonic treatment effect is closely related to the
    quantile treatment effect (QTE) $Q_{Y_1} (t) - Q_{Y_0} (t)$, $t \in [0,1]$,
    whose early formulations date back to Lehmann~\cite{lehmann} and
    Doksum~\cite{doksum}. 
    Importantly for the derivations that follow, however, the comonotonic
    treatment effect is defined as a random variable on the original probability
    space (also contrasting with the previously defined countermonotonic
    treatment effect). 
}
\[
    \Delta^+ = Q_{Y_1} (V_0) - Q_{Y_0} (V_0)
\]
This comonotonic treatment effect is 
consistent with rank invariance and stochastically bounds the feasible $\mte$
functions under rank similarity, analogously to the countermonotonic treatment
effect in \autoref{thm:sharp-set}.
To formally state and prove this result, extend the notation for group
$g$-conditional random vectors to $V_d$ and $\Delta^+$. Then: 

\begin{proposition}[Necessary Conditions on $\mte$ Function, Rank Similarity]
    \label{thm:mte-nec-rank}
    Under \autoref{assn:lism}, and \autoref{assn:rank}a, the $\mte$ function
    satisfies: 
    \begin{align}
        \E [ \mte (U_g)] &= \E [ \Delta_g^+ ] \label{eq:e-rank} \\
        \mte (U_g ) & \succeq_{SSD} \Delta_g^+ \label{eq:ssd-rank}
    \end{align}
    for each group $g \in \{a,c,n\}$. 
\end{proposition}

\noindent
\autoref{thm:mte-nec-rank} is reminiscent of \autoref{thm:sharp-set} but does
not seek to characterize the sharp set of candidate $\mte$ functions under the
additional rank similarity assumption. 
Indeed, if the distribution of each $\Delta_g^+$ is identified from the data, then 
such a characterization follows from arguments analogous to the proof of
\autoref{thm:sharp-set} and the fact that rank similarity does not impose any
further joint restrictions on unobservables and treatment effects within or
across groups $g$. 

Now consider identification of the distributions of group-conditional
comonotonic treatment effects $\Delta_g^+$.  
Given the same data (\autoref{assn:data}), Chernozhukov and
Hansen~\cite{chernozhukovhansen} have derived moment conditions that can
point-identify the distributions of potential outcomes, and thus the
distribution of $\Delta^+$, under rank similarity and an additional assumption
that the potential outcomes are continuous.
To move beyond compliers in the latent index selection model,%
\footnote{
    The comonotonic effect among compliers $\Delta_c^+$ is identified even in the
    basic model (\autoref{assn:lism}) from knowledge of the complier quantile
    functions.
    See Abadie et al.~\cite{abadieangristimbens} for a study of the Local QTE
    among compliers with an emphasis on incorporating covariates. 
}
I introduce an alternative identification strategy that leverages the added
structure of the selection rule \eqref{eq:selection} to accommodate discrete
responses and mass points. 
For example, this is useful for the empirical application of \autoref{sec:ohie}
because most experiment participants never visit the emergency room and thus
receive an outcome of zero. 
The alternative identification strategy, which inherently relies on the
additional structure of the latent index selection model,%
\footnote{
    Or the empirically equivalent monotonicity model of Imbens and
    Angrist~\cite{imbensangrist}. 
} 
is to extrapolate from compliers. 

Extrapolation from compliers consists of three steps. 
First, knowledge of both potential outcome distributions among compliers imposes
restrictions between treated and untreated quantile functions for the
population, and thus for any other subpopulation.   
These restrictions can be summarized in terms of quantile-quantile ($QQ$) plots
of potential outcomes, defined for the population as:   
\begin{equation}
    \label{eq:qq}
    QQ = \{ (Q_{Y_0} (q), Q_{Y_1} (q)): q \in [0,1] \}
\end{equation}
Among subgroups $g$, define an analogous plot $QQ_g$ by replacing population
quantile functions with group-conditional quantile functions. 
The next result establishes the restriction imposed by the complier $QQ_c$ plot
on the population under rank similarity. 

\begin{lemma}
    \label{thm:qq} 
    Under \autoref{assn:lism} and \autoref{assn:rank}a, the conditional
    quantile-quantile plot of any event measurable with respect to $U$ is a
    subset of the population $QQ$ plot. 
    In the case of compliers,
    \[
        Q Q_c \subseteq Q Q.  
    \]
\end{lemma}

\noindent
To simplify the remaining exposition, I also make an additional support
assumption to ensure that the population $QQ$ plot is completely identified from
compliers, i.e. $Q Q_c = Q Q $. 
\begin{assumption}[Full Support Among Compliers] 
    \label{assn:support}
    The complier and population distributions of potential outcomes have equal
    support: $Q_{Y_{d,c}} ([ 0,1]) = Q_{Y_{d}} ([ 0,1]) $. 
\end{assumption}
\noindent
The assumption does not presume or imply continuity of the potential outcomes,
and it is made primarily for technical convenience.%
\footnote{
    The validity of the support assumption is often an empirical question, which
    is answered affirmatively when complier potential outcomes take all possible
    values.  
    For example, in the simplest non-trivial case of binary outcomes, it is easy
    to verify whether both possible realizations sometimes occur.  
}

In the second step, the restrictions from the complier $QQ_c$ plot are combined
with knowledge of one potential outcome distribution among the always- and
never-treated to (partially) recover the other, previously unidentified
potential outcome distribution.
That is, define the complier-extrapolated quantile bounds:
\begin{align}
    \overline{Q}_{Y_{d,g}} (q) 
    &= \sup \{ y_d: (y_0,y_1) \in Q Q_c, \, y_{1-d} = Q_{Y_{1-d,g}} (q) \}
    \label{eq:bq1} \\
    \underline{Q}_{Y_{d,g}} (q) &= \, \inf \{ y_d: (y_0,y_1) \in Q Q_c, \, y_{1-d} = Q_{Y_{1-d,g}} (q) \} 
    \label{eq:bq2}
\end{align}
for all $q \in [0,1]$. Then:  
\begin{lemma}
    \label{thm:extrapolate-quantiles}
    Under \autoref{assn:data}, \autoref{assn:lism}, \autoref{assn:rank}a, and
    \autoref{assn:support}, all quantile functions satisfy the
    complier-extrapolated quantile bounds: 
    \begin{equation}
        \label{eq:extrapolate-quantiles}
        \underline{Q}_{Y_{d,g}} (q)        
        \leq Q_{Y_{d,g}} (q) \leq 
        \overline{Q}_{Y_{d,g}} (q) 
        \quad \text{for all $q \in [0,1]$.}
    \end{equation}
    The bounds are uniformly sharp for $(d,g) \in \{ (0,a), (1,n) \}$, and any
    combination of bounds across groups is jointly attainable by a single
    data-generating process in the population. 
\end{lemma}

\noindent
Under rank similarity (\autoref{assn:rank}a), the identified $QQ_c$ plot serves
as a dictionary for recovering pointwise bounds on the missing quantile
function, using the quantile function identified in the basic model.  
If the potential outcomes are continuously distributed, then the dictionary is
one-to-one, and
all group-conditional potential outcome quantile functions are point-identified.

Finally, in the third step, bounds on the group-conditional comonotonic
treatment effect distributions are obtained under rank invariance, i.e.~by
matching quantiles. 
Fix $V \sim U[0,1]$ and consider random variables with the following
distributions:
\begin{align*}
    \begin{array}{cc} 
    \multicolumn{2}{c}{ 
        \overline{\Delta}_c^+ 
        \stackrel{d}{=} Q_{Y_{1,c}} (V) - Q_{Y_{0,c}} (V) \stackrel{d}{=} 
        \underline{\Delta}_c^+
    } \\[0.1in]
    \overline{\Delta}_a^+ \stackrel{d}{=} Q_{Y_{1,a}} (V) - \underline{Q}_{Y_{0,a}} (V) 
    \quad & \quad 
    \underline{\Delta}_a^+ \stackrel{d}{=} Q_{Y_{1,a}} (V) - \overline{Q}_{Y_{0,a}} (V) 
    \\
    \overline{\Delta}_n^+ \stackrel{d}{=} \overline{Q}_{Y_{1,n}} (V) - Q_{Y_{0,n}} (V) 
    \quad & \quad 
    \underline{\Delta}_n^+ \stackrel{d}{=} \underline{Q}_{Y_{1,n}} (V) - Q_{Y_{0,n}} (V) 
\end{array} 
\end{align*}
In each group, at least one of the quantile functions is identified in the basic
latent index selection model, and uniform bounds on the other quantile function
are obtained from \autoref{thm:extrapolate-quantiles}.
The following first-order dominance relation is immediate by
\eqref{eq:extrapolate-quantiles} of \autoref{thm:extrapolate-quantiles}. 

\begin{lemma}
    \label{thm:rank-te-fsd} 
    Under \autoref{assn:data}, \autoref{assn:lism}, \autoref{assn:rank}a, and
    \autoref{assn:support}, the group-conditional comonotonic treatment effect
    is bounded by:  
    \begin{equation}
        \underline{\Delta}_g^+ 
        \preceq_{FSD} \Delta_g^+ \preceq_{FSD} 
        \overline{\Delta}_g^+ 
        \label{}
    \end{equation}
    The bounds are uniformly sharp for each group and jointly attainable across
    groups by a single data-generating process in the population.  
\end{lemma}

\noindent
The bounds of \autoref{thm:rank-te-fsd} are useful for several purposes. 
First, they undergird an extension of the sharp bounds on $\te (w)$
of \autoref{thm:sharp-te} to the rank similar latent index selection model. 

\begin{proposition}[Bounds on $\te$ Functionals, Rank Similarity]
    \label{thm:sharp-te-rank}
    Under \autoref{assn:data}, \autoref{assn:lism}, \autoref{assn:rank}a, and
    \autoref{assn:support}, 
    sharp bounds on $\te (w)$ for $w (\cdot) \geq 0$ are attained by a pair of
    solutions $\underline{m}_w, \overline{m}_w$ satisfying: 
    \begin{equation}
        \label{eq:te-sharp-rank}
    \begin{array}{c} 
        \overline{m}_w (U_g) \stackrel{d}{=} \overline{\Delta}_g^+ \\[0.05in] 
        \underline{m}_w (U_g) \stackrel{d}{=} \underline{\Delta}_g^+ 
    \end{array} 
    \quad \text{and} \quad
    \begin{array}{l}
        w(U_g) \text{ and } \overline{m}_w (U_g) \text{ are comonotonic} \\[0.05in]
        w(U_g) \text{ and } \underline{m}_w (U_g) \text{ are countermonotonic} 
    \end{array} 
    \end{equation}
    for all $g \in \{a,c,n\}$.
\end{proposition}

\noindent
The bounds on functionals with nonnegative weights are attained at the extremal
identified distributions of comonotonic treatment effects. 
An important feature of \autoref{thm:sharp-te-rank} relative to
\autoref{thm:sharp-te} is that it provides finite bounds on all finite
functionals $\te (w)$ with nonnegative weights, and not just those with nonzero
weights among compliers. 
This arises because rank similarity suffices for true extrapolation from the
compliers whose potential outcomes are collectively identified in the basic
model.  

The bounds of \autoref{thm:rank-te-fsd} can also be combined with
\autoref{thm:mte-nec-rank} to obtain uniformly sharp bounds on counterfactual
averages under rank similarity. 

\begin{proposition}[Uniformly Sharp Bounds on Counterfactual Averages, Rank Similarity] 
    \label{thm:bounds-cao-rank}
    Under \autoref{assn:data}, \autoref{assn:lism}, \autoref{assn:rank}a, and
    \autoref{assn:support}, uniformly sharp bounds on the counterfactual average
    function $\by (p)$ is are given by: 
    \footnotesize
    \[
        \begin{array}{r} 
            \by (\bp_0) - \bp_0 \cdot \left[ 
                \E [ \overline{\Delta}_a^+]
                - I_{\overline{\Delta}_a^+} \left( \frac{p}{\bp_0} \right)
            \right] \\[0.1in]
            \by (\bp_0) + (\bp_1 - \bp_0) \cdot I_{\Delta_c^+}
            \left( \frac{p - \bp_0}{\bp_1 - \bp_0} \right) \\[0.1in]
            \by (\bp_1) + (1 - \bp_1) \cdot I_{\underline{\Delta}_n^+}
            \left( \frac{p - \bp_1}{1 - \bp_1} \right)
        \end{array} 
        \leq \by (p) \leq
        \begin{array}{ll}
            \by (\bp_0) - \bp_0 \cdot I_{\underline{\Delta}_a^+} \left( 
                \frac{\bp_0 - p}{\bp_0}  
                \right) & \text{for $p \in [0,\bp_0]$} \\[0.1in]
                \by (\bp_1) - (\bp_1 - \bp_0) \cdot I_{\Delta_c^+}
                \left( \frac{ \bp_1 - p}{ \bp_1 - \bp_0 } \right) 
                & \text{for $p \in [\bp_0, \bp_1]$} \\[0.1in]
                \by (\bp_1) + (1- \bp_1) \cdot \left[ 
                    \E [ \overline{\Delta}_n^+ ] - 
                    I_{\overline{\Delta}_n^+} \left( \frac{1-p}{1-\bp_1} \right)
                \right] & \text{for $p \in [\bp_1, 1]$}
        \end{array} 
    \]
\end{proposition}

\noindent
In turn, the uniformly sharp bounds impose a necessary condition on the
counterfactual average function.%
\footnote{
    It is worth noting that the uniform sharpness of the bounds of
    \autoref{thm:bounds-cao-rank} relies on the full support condition of
    \autoref{assn:support}. 
    In its absence, the bounding quantile definitions \eqref{eq:bq1} and
    \eqref{eq:bq2} must be generalized, and then the extremal comonotonic
    effects underlying the lower or upper bound, e.g. $\overline{\Delta}_a^+$
    and $\underline{\Delta}_n^+$, may no longer be \emph{jointly} attainable by
    a single rank similar data-generating process.  
    Nevertheless, pointwise sharpness suffices for the necessary condition of
    interest.  
    The issue also does not arise in \autoref{thm:sharp-te-rank} because the
    underlying extremal comonotonic effects in that case, e.g.
    $\overline{\Delta}_a^+$ and $\overline{\Delta}_n^+$ for the upper bound,
    remain jointly attainable.
}
This yields a simple test of whether rank similarity and further assumptions
within the latent index selection framework are jointly consistent. 
Thus, for example, I find suggestive evidence that rank similarity is not
consistent with the linearity assumption of Brinch et al.~\cite{brinch} in the
empirical application of \autoref{sec:ohie}. 
In other words, the linear extrapolation could not be generated by a rank
similar or invariant process, even though such a distribution assumption may be
reasonable in the context of health expenditures.  

While rank similarity may be inconsistent with ancillary assumptions, it is
always consistent with the basic latent index selection model itself. 
Intuitively, this follows from the logic of extrapolating from compliers in
\autoref{thm:qq}. 
Any marginal distributions of complier outcomes define a rank invariant joint
distribution of complier outcomes through the operation of matching quantiles. 
Any such joint distribution of complier outcomes can be extended to the always-
and never-treated because in each case one outcome is unobserved and therefore
unconstrained.
If \autoref{assn:support} is violated and the supports of the observed
outcome distributions vary across the never/always-treated and compliers, then
the joint distribution of population outcomes may simply not be sufficiently
identified to allow meaningful extrapolation without further assumptions. 

Rank similarity can also be imposed conditional on observed covariates $X$
(along with \autoref{assn:lism}). 
It is useful to distinguish between two formulations. 

\begin{assumption}[Conditional Rank Similarity] 
    \label{assn:rank-cov}
    Letting $Y_{d,x} \sim (Y_d | X=x)$, 
    \begin{enumerate}
        \item[a.] Potential outcomes satisfy weak conditional rank similarity if for
            $d=0,1$ there exist rank variables $\tilde{V}_d$ that satisfy $Y_d =
            Q_{Y_{d,X}} (\tilde{V}_d) $ and that are identically distributed conditional
            on $(U,X)$. 
        \item[b.] Potential outcomes satisfy strong conditional rank similarity
            if for $d=0,1$ there exist rank variables $V_d$ that satisfy $Y_d =
            Q_{Y_d} (V_d)$ and that are identically distributed conditional on
            $(U,X)$.  
    \end{enumerate}
\end{assumption}

\noindent
The first version of conditional rank similarity (\autoref{assn:rank-cov}a)
imposes the previous unconditional rank similarity (\autoref{assn:rank}a) within
each subpopulation $X=x$.  
As in the preceding extension (\autoref{sec:covars}), the same analysis then
proceeds within subpopulations. 
Aggregating across subpopulations, define (on the original probability space) the
covariate-conditional comonotonic effect by: 
\[
    \Delta_{\bar{x}}^+ = Q_{Y_{1,X}} (\tilde{V}_0) - Q_{Y_{0,X}} (\tilde{V}_0)
\]
Under \autoref{assn:policies}, the group- and covariate-conditional comonotonic
effect $\Delta_{g,\bar{x}}^+ \sim (\Delta_{\bar{x}}^+ | G=g)$ constrains the set
of feasible $\mprte$ functions, like the group-conditional comonotonic effect
$\Delta_g^+$ constrained the $\mte$ function in \autoref{thm:mte-nec-rank}. 
However, whereas covariate information sharpened bounds in the basic model,
weak conditional rank similarity relaxes bounds relative to the unconditional
rank similarity assumption because the covariate-conditional comonotonic effect
is feasible in the basic model and therefore invoking \eqref{eq:te-ssd}
implies: 
\begin{equation}
    \label{eq:rank-ssd}
    \Delta_{g,\bar{x}}^+ \preceq_{SSD} \Delta_g^+ \quad \text{for all $g \in
    \{\emptyset, a,c,n\}$}
\end{equation}
where $g = \emptyset$ is used to denote the unconditional case. 

The second version of conditional rank similarity (\autoref{assn:rank-cov}b) is
theoretically stronger than unconditional rank similarity (\autoref{assn:rank}a)
because it conditions equality of rank distributions on the relatively finer
realizations of $(X,U)$.  
However, because the treatment effect $\Delta^+$ satisfying unconditional rank
invariance (\autoref{assn:rank}b) is feasible and extremal in either case, the
second version of conditional rank similarity confers no additional identifying
power relative to unconditional rank similarity in practice.  
As the name suggests, the second version of conditional rank similarity is also
stronger than the first.
Relatedly, strong conditional rank similarity (\autoref{assn:rank-cov}b) is
testable with information on covariates [Dong and Shen~\cite{dongshen}, Frandsen
and Lefgren~\cite{frandsenlefgren2018}]. 
The present framework suggests two new, albeit related, testable implications.  

\begin{proposition}[Conditional Rank Similarity]  
    \label{thm:cond-rs}
    The strong version of conditional rank similarity (\autoref{assn:rank-cov}b)
    implies the weak version (\autoref{assn:rank-cov}a).  
    Furthermore, under \autoref{assn:lism}, the strong version implies the restrictions: 
    \begin{equation}
        \label{eq:strong1}
        QQ_{g,x} \subseteq QQ 
    \end{equation}
    and: 
    \begin{equation}
        \label{eq:rank-ssd-strong}
        \Delta_{g,\bar{x}}^+ \stackrel{d}{=} \Delta_g^+. 
    \end{equation}
    for groups $g = \{ \emptyset, a,c,n \}$ and covariate realizations $x$. 
\end{proposition}
    
\noindent
The first implication \eqref{eq:strong1} requires that the conditional
$QQ_{c,x}$ plots across realizations $x$ must lie on a single totally ordered
curve in $\mathds{R}^2$, which is testable with identified covariate-conditional
quantiles, e.g.~among compliers. 
This is essentially the main testable implication of Dong and Shen~\cite{dongshen}
applied to potential outcomes rather than the underlying ranks.%
\footnote{
    The main testable implication of Dong and Shen~\cite{dongshen} is that the
    distribution of ranks is equal across treatment status conditional on
    covariates. 
    This implies the $QQ$ plot relation, and the reverse implication is also
    true under their assumption that potential outcomes are continuous.  
    An advantage to an outcome test is that the quantile-quantile plots are
    unique even when the rank variables are not, as is the case when potential
    outcomes are not continuous. 
}
The second implication \eqref{eq:rank-ssd-strong} requires that
\eqref{eq:rank-ssd} holds with equality in distribution.  
This implies equality of the uniform bounds among compliers under weak
conditional rank similarity and unconditional rank similarity. 
Thus \eqref{eq:rank-ssd-strong} can be visually assessed in the graphical plot
suggested in \autoref{sec:sum-bounds} and implemented empirically in
\autoref{sec:ohie}. 
I now turn to this empirical application. 

\section{Empirical Application}
\label{sec:ohie}

This section applies the methods and framework of the previous sections using
publicly available data from the Oregon Health Insurance Experiment (OHIE)
[Finkelstein~\cite{finkelstein-data}]. 
In 2008, a group of uninsured, low-income adults in Oregon were offered the
chance to apply for Medicaid via a random lottery. 
This experiment has provided a unique opportunity to identify
causal effects of health insurance across a variety of outcomes among lottery
compliers.  
At the same time, using the data to inform many questions of policy interest ---
such as the potential effects of charging a price for similar coverage, or of
proposed expansionary policies like ``Medicare for All'' --- requires
interpolating or extrapolating beyond point-identified local average treatment
effects.  
Thus the setting is appealing for employing the methods proposed in this paper. 

Specifically, I contribute to the study of the effects of Medicaid on emergency
room (ER) utilization. 
Emergency rooms are an expensive and, when used for non-emergencies,
inefficient form of medical care. 
Furthermore, the effect of expanded medical insurance coverage on emergency room
utilization is theoretically ambiguous: health insurance simultaneously
decreases costs of both emergency and non-emergency care, as well as possibly
affecting health outcomes directly. 
In summary, the impact of Medicaid expansion on emergency care is an important
empirical policy question. 

The effect of Medicaid on emergency room utilization has been previously studied
in the randomized control setting of the OHIE by Taubman et al.~\cite{taubman}
and Kowalski~\cite{kowalski2016, kowalski2021}. 
Taubman et al.~\cite{taubman} find positive causal effects of Medicaid
enrollment on the number of emergency room visits. 
These effects are significant (on the order of 40\% relative to a control group)
across a wide range of visit types, conditions, and subgroups.
Kowalski~\cite{kowalski2016} extends this analysis by assessing the external
validity of the $\late$ and conducting a variety of policy-relevant
extrapolation exercises for three ER utilization outcomes --- whether a
participant visited the ER, the number of ER visits, and the number of total
ER charges --- under the additional assumptions of marginal outcome monotonicity
and linearity. 
Kowalski~\cite{kowalski2021} further extends the analysis, with a focus on
reconciling the results with those from a previous 2006 health reform in
Massachusetts. 
The following paragraph discusses these previous findings in more detail, in
order to compare to my own.

Kowalski~\cite{kowalski2016} finds that, under marginal outcome monotonicity,
the data is consistent with an externally valid LATE for all ER
utilization outcomes. 
Under the stronger linearity assumption of Brinch et al.~\cite{brinch}, the
$\mte$ function is point-identified and decreasing in the propensity for
treatment, indicating a higher expected treatment effect for those more likely
to select into insurance. 
For all ER utilization outcomes, $\mte (u)$ is positive for all always-takers
and negative for all or most never-takers. 
Furthermore, the linear extrapolation predicts that further expanding Medicaid
coverage to the never-treated (who were either ineligible or would have chosen
not to enroll in the experimental Medicaid expansion) would generate negative
local average treatment effects among that subpopulation.  
This is in contrast to the positive average effect identified among compliers
(who became eligible and enrolled in Medicaid under the expansion), and the even
larger positive treatment effects predicted among the always-treated (who were
already eligible for Medicaid).%
\footnote{
Kowalski~\cite{kowalski2021}
uses this extrapolation to reconcile the OHIE results with the seemingly
contradictory results from a 2006 Massachusetts health reform, which showed that
ER utilization decreased or stayed the same (Chen et al.~\cite{chen}, Smulowitz
et al.~\cite{smulowitz}, Kolstad and Kowalski~\cite{kolstadkowalski},
Miller~\cite{miller}). 
Specifically, Kowalski~\cite{kowalski2021} argues that Massachusetts compliers
are similar to Oregon never-takers, for whom the linear latent index model
implies negative treatment effects.
}
Thus the linear extrapolation has significant policy implications.  
Finally, 
there is some evidence that the functional form is misspecified in the case of
ER charges: linearity implies that marginal treatment outcomes are negative for
45\% of the sample, which is impossible because ER charges are nonnegative by
nature.   

The present analysis builds on the preceding work as follows.%
\footnote{
The publicly available OHIE data consists of 24,646 observations corresponding
to lottery entrants who lived in a postal code where residents almost
exclusively used one of the twelve hospitals for which hospital usage data was
collected. 
Following the analysis of Kowalski~\cite{kowalski2016}, I further restrict to
the subsample of 19,643 entrants who were the only entrants in their households,
in order to maintain internal validity.  
    It should be noted that variables in the public-use data have been censored and
    truncated in order to limit the identification of participants; 
    however, the analysis of Kowalski~\cite{kowalski2016} suggests that this should
    not significantly affect the conclusions reached.  
Since I study the same ER outcomes for the same subsample as
Kowalski~\cite{kowalski2016}, the reader is referred to her empirical analysis
for a set of summary statistics and a detailed study of outcomes and observable
characteristics across different treatment groups.  
    Finally, to maintain the focus on identification, standard errors are
    omitted, and the following statements are based on point estimates and not
    statistical significance.  
}
First, incorporating distributions rather than just means yields sharp
interpolation bounds in the basic latent index selection model, such as for the
$\late (\bp_0, \bp_1 - 0.1 (\bp_1 - \bp_0))$ corresponding to a 10\% reduction
of compliers (bounds for all treatment effects referenced in the text are
provided in \autoref{tab:lates}).
Such interpolations are policy-relevant; for example, as discussed by
Kowalski~\cite{kowalski2016}, a policy maker may consider providing lottery
winners the opportunity to receive discounted rather than free insurance
coverage, inducing some experimental compliers to decline coverage. 

For example, for the number of ER visits, the model implies that this
aforementioned LATE with discounted coverage is bounded below by -0.66 and above
by 1.11 additional ER visits due to insurance coverage.
The width of the bounds reflects
the fact that the model places little structure on i) how potential outcomes are
jointly distributed among compliers, and ii) which compliers are most likely to
decline coverage if it is discounted rather than free.
Thus, removing 10\% of compliers could have a variety of effects that remain
consistent with the basic model.  
More positively, the interpolations are finite without further assumptions, and
they typically improve with covariate information, which constrains feasible extremal
distributions. 
For example, including covariate information tightens the bounds to
[-0.53,0.98].
Thus, for example, we can conclude that while discounted coverage leading to a
10\% reduction in compliers could reduce ER visits among the corresponding
compliers, it would not do so by more than roughly half a visit on average. 

In comparison, the existing mean-consistent method of Mogstad et
al.~\cite{mogstad} only recovers the sharp bounds for the binary ER visit
outcome, because binary
outcome distributions are summarized by their means. 
For the other outcomes, some interpolation from means remains possible 
because potential outcomes are bounded below, $Y_d \geq 0$; otherwise there
exist mean-consistent distributions with arbitrary dispersion, and so no finite
interpolation is possible using existing methods without further assumptions. 
Even with nonnegativity, the mean-consistent bounds of [-1.30,1.60] on the
aforementioned $\late$ with discounted coverage are notably wider than those
that use additional information from identified distributions.
For example, they fail to rule out that the discounted coverage will reduce ER
visits among the corresponding compliers by at least one on average, even though
such inference is possible given the assumptions of the model. 

As a second contribution, incorporating distributions allows one to incorporate
further distribution assumptions, such as (conditional) rank similarity, into
the latent index selection model.   
In the OHIE context, rank similarity is satisfied if those who have relatively
higher ER utilization without insurance also have relatively higher ER
utilization with insurance.    
In other words, receipt of insurance does not systematically change the ordering
of utilization in the population. 
A consistent finding across ER utilization outcomes under (conditional) rank
similarity is that extrapolations to the never-treated yield a nonnegative
counterfactual average treatment effect, $\late(\bp_1, 1)$.%
\footnote{
    Methodologically, the proposed identification method leverages the added
    structure of the latent index selection model to (partially) identify
    distributions even though two of the ER outcomes are discrete and all
    ER outcomes have a mass point at zero.        
    Because the outcomes are thus discontinuous, the previous identification
    result of Chernozhukov and Hansen~\cite{chernozhukovhansen} does not apply. 
}
This presents a stark contrast to the linear extrapolation, which predicts a
\emph{negative} average treatment effect among the never-treated.     
For the indicator and number of ER visit outcomes, unconditional rank similarity
makes the even stronger prediction that the $\mte$ function is everywhere
positive, because the complier quantile functions are consistently ranked on
their domain.
Thus partial extrapolations, such as the $\late (\bp_1, \bp_1 + 0.1(\bp_1 -
\bp_0))$ corresponding to a 10\% expansion of compliers, are also bounded below
by zero for the two discrete ER outcomes.  

For the total ER charges outcome, the fact that complier quantile functions are
not ranked implies that the realized treatment effects among some participants
are negative;%
\footnote{
    In other words, $P ( Y_1 < Y_0 | G = c) > 0$. 
    For a detailed discussion of bounds on such ``inequality probabilities'' in
    the context of health economics, see Mullahy~\cite{mullahy2018}.
}
thus some counterfactual average effects could also be negative.
Yet the extrapolated effect $\late(\bp_1,1)$ among the never-treated is even
larger than the $\late_{IA} = \late(\bp_0, \bp_1)$ among compliers but smaller
than the extrapolated effect $\late(0, \bp_0)$ among the always-treated. 
This is inconsistent with the weaker $\mte$ monotonicity assumption.
The non-monotonicity occurs because (conditional) rank similarity predicts that
experiment participants in the extreme upper tail of the total charge
distribution have large positive treatment effects, which account for most of
the mean effect, and these participants are more highly represented among the
never-treated than among compliers.  
This finding remains consistent with the adverse selection on health observed by
Kowalski~\cite{kowalski2021} because i)   
adverse selection in health across treatment propensity is distinct from
variation in marginal treatment effects (or moral hazard) across treatment
propensity, and ii) the finding is driven by the extreme right tail of
expenditures among the never-treated, and not a discrepant ranking of mean
untreated outcomes. 
Therefore, while estimated averages of distributions with such high dispersion
are noisy, the estimates still suggest a plausible extrapolation story that
differs from the one obtained by functional form assumptions. 
Summarizing the main novel empirical takeaway, I find that full extrapolations
to the never-treated under rank similarity yield a consistently nonnegative
treatment effect across a variety of measures of utilization, in contrast to
previous findings of a negative treatment effect under a linear extrapolation of
the $\mte$ function.

\begin{figure}[H]

    \vspace{-1.5in}
\centering
\includegraphics[width=\linewidth]{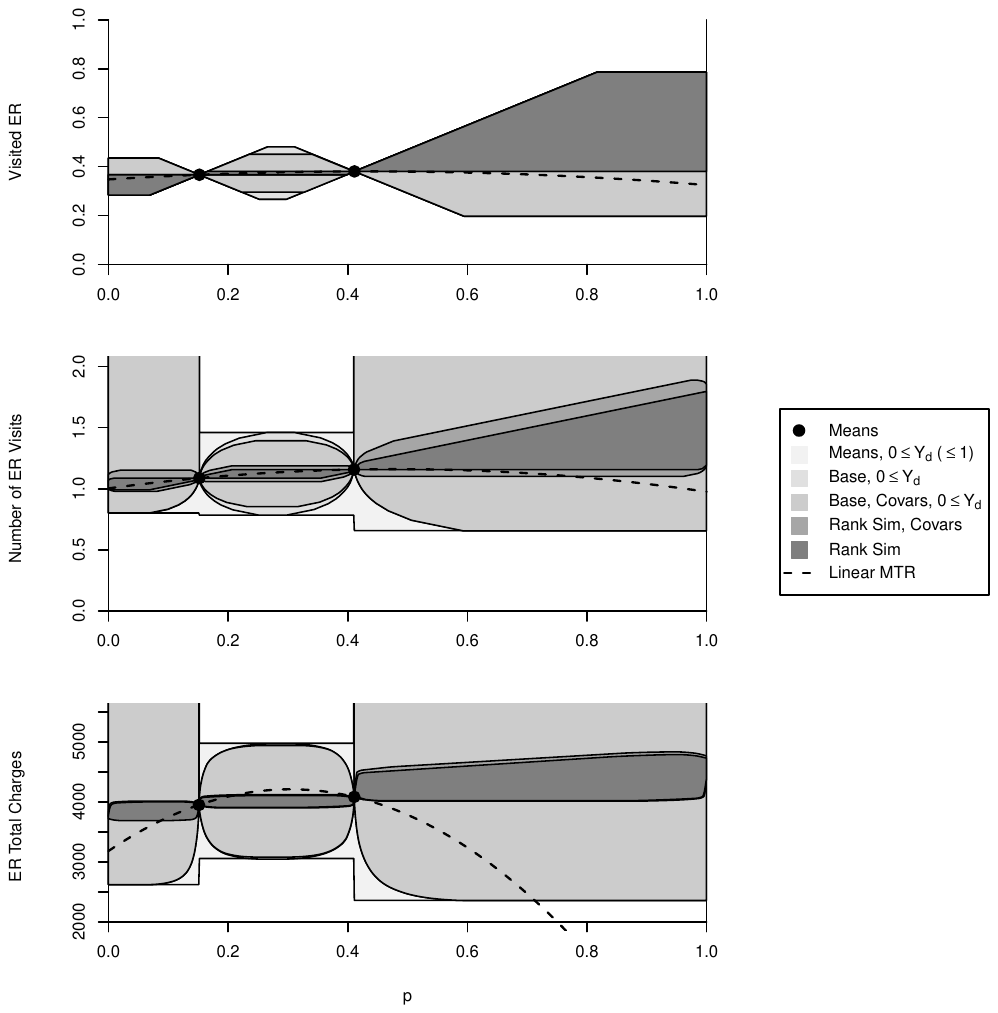} 

\vspace{-1.5in}
\caption{
    \label{fig:ubounds}
    Uniformly Sharp Bounds on Counterfactual Average Outcomes
    for OHIE Data.
    This figure empirically implements the uniformly sharp bounds on
    counterfactual average outcomes of \autoref{thm:sharp-avg} for the OHIE data
    across the three ER utilization outcomes: whether a participant visited the
    ER, the number of ER visits, and the total ER charges incurred. 
    These uniform bounds summarize the empirical content of the (rank similar
    and/or covariate-conditional) latent index models for the family of
    (positively) weighted average treatment effects. 
    In particular, bounds for the counterfactual $\late$ parameters in
    \autoref{tab:lates} could be derived from the information in this figure.  
    In the case of the binary indicator of whether an experiment subject visited
    the ER, the mean and distribution bounds coincide because distributions are
    characterized by their means. 
    In the other cases, incorporating distribution information improves
    identification, and also allows the researcher to incorporate distribution
    assumptions, such as rank similarity. 
    The uniform bounds illustrate how the extrapolations based
    on rank similarity differ substantively from the parametric extrapolation
    based on means alone. 
}
\end{figure}

At the same time, the aim of the preceding comparison is not to argue that rank
similarity is necessarily a better assumption than outcome monotonicity or
linearity in the OHIE context, but rather to highlight the third contribution:
the framework and its incorporation of distribution information allows the
researcher to compare the content of various mean and distribution assumptions
and assess their joint consistency.  
As an additional such example, in \autoref{fig:ubounds}, the $\mte$ function
obtained under the linearity assumption for total charges is too dispersed to
have been generated by a (conditionally) rank similar data-generating process
among compliers.   
Thus again, the linearity and rank similarity assumptions do not appear jointly
consistent.  
More generally, this shows how distribution information alone could in theory
reject functional-form assumptions that point-identify functional parameters of
interest, as in \autoref{thm:linear-specification}. 
Additionally, the uniform bounds in \autoref{fig:ubounds} provide a
specification test of unconditional rank similarity, as formalized in
\autoref{thm:cond-rs}.  
In this case, the visible difference in unconditional and conditional rank
similar bounds for the number of ER visits provides some suggestive evidence
against unconditional rank similarity for that outcome;
however, such evidence is nonexistent for the binary visit outcome and seemingly
more limited for the ER total charge outcome.     
The statistical formalization of these observations is left to future work.  

Finally, the preceding insights --- as well as the features of data required for
sharply bounding all weighted treatment effects $\te (w)$ across the different
auxiliary assumptions --- are compactly summarized in the uniform bounds of
\autoref{fig:ubounds}.  
Thus the proposed methods provide an overarching framework for the partial
identification of treatment effects in the latent index selection model.  

\section{Conclusion} 
\label{sec:conclusion}

This paper characterizes the set of feasible $\mte$ functions and derives sharp,
analytic bounds on weighted average treatment effects in the latent index
selection model [Heckman and Vytlacil~\cite{hv1999, hv2005};
Vytlacil~\cite{vytlacil}]. 
The latent index selection model is a popular instrumental-variable framework
for identification, which in many empirical settings defines a larger class of
parameters than it identifies.  
For example, in the empirical setting of the Oregon Health Insurance Experiment,
the researcher or policymaker may be interested in the effects of charging a
price for expanded insurance coverage, or of expanding coverage to an even
larger population.
A theme of the paper is that sharp characterizations of the latent index
selection model's empirical content fully utilize identified marginal
distributions. 
Furthermore, such distribution information is useful for encoding new auxiliary
assumptions into the latent index framework, which may yield quite different
predictions than assumptions that use means alone.     

There are several interesting possibilities for future work. 
One is to (sharply) incorporate other auxiliary distribution
assumptions into the model. 
Another is to study whether the analytical expressions for bounds can be used to
simplify results about statistical inference. 
Finally, and most generally, it would be interesting to study how the analytical
toolkit and results in this paper could be used to obtain sharp, analytical
results in generalized models that relax the binary nature of the treatment
decision or the structure of the latent index selection rule. 

\appendix

\section{Empirical Table and Figure} 

\input{empirical/tables/wte-bounds.tex}

\section{Proofs} 

\begin{proof}[Proof of \autoref{thm:sharp-set}]
    To establish necessity, observe that the definition of the $\mte$ function
    and the group-conditional random vector implies that $\mte (U_g) = \E [
    \Delta_g | U_g] $ for all $g \in \{a,c,n\}$. 
    Then mean consistency \eqref{eq:mean} follows from the law of total
    expectation:
    \[
        \E [ \mte (U_g) ] =
        \E [ \E [ \Delta_g | U_g ]] 
        = \E [ \Delta_g ] . 
    \]
    For $g = c$, the last term is equal to $\late_{IA}$. 
    For distribution consistency,
    Jensen's inequality and the law of total expectation imply: 
    \begin{align*}
        \E [ \phi ( \mte (U_g) ) ] &= \E [ \phi ( \E [ \Delta_g | U_g ] ) ] \\
        &\geq \E [ \E [ \phi (\Delta_g ) | U_g ]] \\
        &= \E [ \phi (\Delta_g ) ] 
    \end{align*}
    for any concave $\phi : \mathds{R} \to \mathds{R}$. 
    Thus $\mte (U_g) \succeq_{SSD} \Delta_g$ follows by an equivalent and common
    characterization of SSD dominance, namely (weakly) higher expectations over
    all increasing concave functions. 
    Additionally, 
    \[
        I_{\Delta_g} (q) \geq \int_{0}^{q} [ Q_{Y_{1,g}} (s) - Q_{Y_{0,g}} (1-s) ]
        \, ds = I_{\Delta_g^-} (q) 
        \quad \text{for all $q \in [0,1]$.}
    \]
    The inequality follows by definition of the treatment effect, $\Delta_g =
    Y_{1,g} - Y_{0,g}$, whose integral of the lowest $q$ values is at least as
    high as the 
    lowest $q$ values of $Y_{1,g}$ and the highest $q$ values of $Y_{0,g}$; the
    equality follows by definition \eqref{eq:Qneg} of $\Delta_g^-$ and the IQF.
    Thus $\Delta_g \succeq_{SSD} \Delta_g^- $ by definition.
    Since the SSD order is transitive, combining results yields $\mte (U_g)
    \succeq_{SSD} \Delta_g^-$. 
    This is an empirical restriction for compliers $g = c$ since the
    distribution of $\Delta_c^-$ is recovered from the identified marginals of
    $Y_{0,c}$ and $Y_{1,c}$. 

    In contrast, no meaningful restrictions on the $\mte$ function are possible
    for non-compliers $g \neq c$ because either the conditional distribution of
    treated or untreated outcomes is unidentified (\autoref{thm:group-lemma}). 
    By definition of the $\mte$ function \eqref{eq:mte} and by
    \autoref{thm:group-lemma}, the sharp set $\M^*$ is equal to the set of
    integrable functions $m$ that satisfy: 
    \begin{equation}
        m (\hat{U}_g) = \E [ \hat{Y}_{1,g} - \hat{Y}_{0,g} | \hat{U}_g ]  
        \label{eq:mte-group}
    \end{equation}
    for some random vector $ (\hat{U}_g, \hat{Y}_{0,g}, \hat{Y}_{1,g})$
    consistent with identified marginals, for every $g \in \{a,c,n\}$. 
    However, for any integrable function $m$ and, say, the always-treated
    $g = n$, we can choose a conditional joint distribution of $(\hat{U}_n,
    \hat{Y}_{0,n}, \hat{Y}_{1,n})$ that is consistent with the identified
    marginals of $U_n$ and $Y_{0,n}$ and that satisfies \eqref{eq:mte-group};
    namely, choose the unidentified treated outcome distribution among the
    never-treated to satisfy $\E [ \hat{Y}_{1,n} | \hat{U}_n] = m (\hat{U}_n) +
    \E [ \hat{Y}_{0,n} | \hat{U}_n ]$ for any data-consistent $(\hat{U}_n,
    \hat{Y}_{0,n})$. 
    Analogous reasoning excludes the possibility of restricting the $\mte$
    function on the interval of the always-treated. 

    To establish sufficiency, it therefore remains to show that if an integrable
    function $m$ satisfies \eqref{eq:mean} and \eqref{eq:ssd}, there exists a
    random vector $ (\hat{U}_c, \hat{\Delta}_c = \hat{Y}_{1,c} - \hat{Y}_{0,c})$
    consistent with the identified marginals of $U_c$,
    $Y_{0,c}$, and $Y_{1,c}$ such that \eqref{eq:mte-group} holds. 
    Conditions \eqref{eq:mean} and \eqref{eq:ssd} are jointly equivalent to
    (reverse) stochastic dominance in the \emph{convex order}:%
    \footnote{
        The convex order is defined as $X \succeq_{cx} X'$ if $\E [ \phi (X) ]
        \geq \E [ \phi (X') ]$ for all convex functions $\phi: \mathds{R} \to
        \mathds{R}$. 
        A reference is Shaked and Shanthikumar~\cite{shakedshanthikumar}; the
        invoked equivalences are given in Theorems 3.A.4 and 3.A.5.  
    }
    \begin{equation}
        m (U_c) \preceq_{cx} \Delta_c^-. 
        \label{eq:mte-cx}
    \end{equation}
    In turn, dominance in the convex order is equivalent to the existence of
    random variables $ \hat{M}_c $ and $\hat{\Delta}_c^-$, defined on the same
    probability space and with marginal distributions equal to those of their
    namesakes, such that the random vector $(\hat{M}_c, \hat{\Delta}_c^-)$ is a
    martingale:%
    \footnote{
        See M\"uller and R\"uschendorf~\cite{mullerruschendorf}, Theorem 4.1,
        for a statement of this equivalence, commonly referred to as Strassen's
        Theorem, as well as an explicit algorithmic construction of the
        corresponding Markov kernels. 
        For a closely related construction in economics, see Machina and
        Pratt~\cite{machinapratt}.  
    }
    \[
        \E [ \hat{\Delta}_c^- | \hat{M}_c ] = \hat{M}_c. 
    \]
    In other words, letting $\mathcal{B}$ denote the Borel $\sigma$-algebra on
    $\mathds{R}$, dominance in the convex order guarantees the existence of a
    Markov kernel $\kappa: \mathcal{B} \times \mathds{R} \to [0,1]$ with:
    \[
        \int t \kappa (dt,x) = x \quad \text{for all $x \in \mathds{R}$}  
    \]
    and $P_{\Delta_c^-} (B) = \int \kappa (B,x) \, P_{m (U_c)} (dx)$ for all $B
    \in \mathcal{B}$. 
    But then the joint distribution generated by a $\hat{U}_c$ with the marginal
    distribution of $U_c$ and the conditional law $[\hat{\Delta}_c^- | \hat{U}_c
    = u]$ corresponding to $\kappa (\cdot, m(u))$ yields a joint distribution
    with empirically consistent marginals satisfying: 
    \[
        \E [ \hat{\Delta}_c^- | \hat{U}_c ] = m (\hat{U}_c). 
    \]
    This is the desired result. 

    Finally, to show that the sharp set $\M^*$ is convex, consider elements
    $m,m' \in \M^*$ with corresponding complier kernels $\kappa, \kappa'$.  
    Then for any $\alpha \in [0,1]$, the joint distribution generated by a
    $\tilde{U}_c$ with the marginal distribution of $U_c$ and the conditional
    law $[ \tilde{\Delta}_c^- | \tilde{U}_c = u]$ corresponding to the mixture
    $\alpha \kappa (\cdot, m(u)) + (1-\alpha) \kappa' (\cdot, m'(u))$  yields a
    joint distribution with empirically consistent marginals satisfying: 
    \[
        \E [ \tilde{\Delta}_c^- | \tilde{U}_c ] = \alpha m (\tilde{U}_c) +
        (1-\alpha)  m' ( \tilde{U}_c )  
    \]
    Since $\alpha m + (1-\alpha) m'$ also inherits integrability, it follows
    that the function belongs to $\M^*$. 

\end{proof}

\begin{proof}[Proof of \autoref{thm:sharp-te}]
    In the basic formulation (\autoref{assn:data} and \autoref{assn:lism}), a
    necessary condition for finite bounds on WTE$(w)$ is that weights are
    nonzero only among compliers:
    \begin{equation}
        \label{eq:nonneg-compliers}
        P (w(U) \neq 0, G \neq c) = 0.
    \end{equation}
    Else, suppose $P (w(U) > 0, G \neq c) > 0$, and let $\U_{-c,+} = \{u: w(u) >
    0, u \notin [\bp_0, \bp_1] \}$.
    (The case where $w$ is instead negative follows analogously.)
    For any candidate finite upper (analogously, lower) bound $k \in
    \mathds{R}$ we can construct a violation as follows.  
    Consider the family of candidate $\mte$ functions:
    \[
        m_n (u) = \begin{cases}
            n                   & \text{if $u \in \U_{-c,+}$} \\
            \late(\bp_0,\bp_1)  & \text{if $u \notin \U_{-c,+}$ }
        \end{cases} 
    \]
    which by Theorem 1 belongs to $\M^*$ for any $n \in \mathds{N}$. 
    Yet for:  
    \[
        n_k > \frac{
            k - \late(\bp_0, \bp_1) E [ w(U) \mathds{I}\{U \notin \U_{-c,+} \}]
        }{
            E [ w(U) \I \{ U \in  \U_{+,-c} \}] 
        }
    \]
    we have that $\int m_{n_k} (u) w(u) \, du > k$, yielding a violation of the
    supposed bound. 
    Thus, no finite bounds on such $\te(w)$ exist, and it suffices to consider
    $\te (w)$ where \eqref{eq:nonneg-compliers} holds and to specify functions
    that extremize $\te (w)$ on the interval of compliers. 

    Fix $w$ and consider an $\overline{m}_w$ satisfying \eqref{eq:functional}.  
    Fix any other $m \in \M^*$, and to simplify notation let $M = m(U_c)$, $W =
    w(U_c)$, and $\bar{M} = \overline{m}_w (U_c)$, and $V \sim U[0,1]$.
    Then: 
    \begin{align}
        \frac{1}{P (G = c)}
        \int_{\bp_0}^{\bp_1} m(u) w(u) \, du 
        &= \E [ M W ] \label{eq:rv} \\
        &\leq \E [ Q_M (V) Q_W (V) ] \label{eq:sm} \\
        &\leq \E [ Q_{\Delta_c^-} (V) Q_W (V) ] \label{eq:dil} \\
        &= \E [ \bar{M} W ] 
        = \frac{1}{P (G=c)} \int_{\bp_0}^{\bp_1} \overline{m}_w (u) w(u) \, du 
        \label{eq:mte-bound}
    \end{align}
    The equality in \eqref{eq:rv} follows by definition of the random variables.  
    The inequality \eqref{eq:sm} follows from Theorem 9.A.21 of Shaked and
    Shantikumar~\cite{shakedshanthikumar} and the fact that $f(x,y) = xy$ is a
    supermodular function.
    Note that $(Q_M (V), Q_W (V))$ is the joint distribution where $M$ and $W$
    are comonotonic, i.e. exhibit perfect positive dependence.%
    \footnote{
        Alternatively expressed, \eqref{eq:sm} also follows from the existence
        of a comonotonic solution to the primal Monge-Kantorovich problem for
        supermodular functions, e.g.  Galichon~\cite{galichon}, Theorem 4.3. 
    }
    Inequality \eqref{eq:dil} follows from Theorem 3.A.9 of Shaked and
    Shantikumar~\cite{shakedshanthikumar}, the sharp set characterization of
    \autoref{thm:sharp-set}, and the
    fact that $Q_W (\cdot) $ is a nondecreasing function. 
    Finally, equality \eqref{eq:mte-bound} follows from the assumed properties
    in \eqref{eq:functional}, which impose comonotonicity between the extremal
    $\mte$ function and the weights.  
    An analogous argument holds for the function $\underline{m}_w$ attaining the
    lower bound by working with the negative weight $-w$. 
\end{proof}

\begin{proof}[Proof of \autoref{thm:sharp-avg}]
    Combining \eqref{eq:int-mte} with the identification of $\by (\bp_z)=\E [ Y
    | Z=z]$,  
    a function $a:[0,1] \to \mathds{R}$ is a counterfactual outcome candidate
    function if and only if it has a representation: 
    \begin{equation}
        \label{eq:a-rep}
        a (p) = \by (\bp_0) + \int_{\bp_0}^p m (u) \, du 
    \end{equation}
    for some marginal treatment effect candidate $m \in \M^*$. 
    Thus any candidate outcome function $a$ is absolutely continuous, with $a
    (\bp_0) = \by (\bp_0)$ and derivative $a ' = m$ almost everywhere, i.e.  $a'
    \in \M^*$. 
    I proceed to show necessity and sufficiency of \eqref{eq:sharp-cao}. 

    \emph{Necessity:} The upper bound of \eqref{eq:sharp-cao} must be satisfied
    by virtue of a decreasing rearrangement and the pointwise upper bound of
    \autoref{thm:bounds-cao}:
    \[
        a (\bp_0) + \int_{\bp_0}^p (a')^{\uparrow_c} (u) \, du 
        \leq 
        a (p) \leq \psi_h (p)  
    \]
    For the lower bound, restricting for simplicity of notation to the complier
    interval $p \in [\bp_0,\bp_1]$ where the pointwise bounds are meaningful,
    \begin{align*}
        a (\bp_0) + \int_{\bp_0}^p (a')^{\uparrow_c} (u) \, du 
        &= a (\bp_0) + \int_{\bp_0}^p Q_{a' (U_c)} 
            \left( \frac{u - \bp_0}{\bp_1 - \bp_0} \right) \, du \\
        &= a (\bp_0) + (\bp_1 - \bp_0) \cdot I_{a' (U_c)} 
            \left( \frac{p - \bp_0}{\bp_1 - \bp_0} \right) \\
        &\geq  \by (\bp_0) + (\bp_1 - \bp_0) \cdot I_{\Delta_c^-} 
            \left( \frac{p - \bp_0}{\bp_1 - \bp_0} \right) = \psi_l (p). 
    \end{align*}
    The inequality follows because $a (\bp_0) = \by (\bp_0)$ and $a' \in \M^*$,
    and therefore $I_{a' (U_c)} (q) \geq I_{\Delta_c^-} (q) $ pointwise by
    \eqref{eq:ssd} and the definition of SSD. 
       
    \emph{Sufficiency:} Fix an absolutely continuous $a:[0,1] \to
    \mathds{R}$, which by definition is differentiable almost everywhere and has
    a representation:
    \[
        a (p) = a (\bp_0) + \int_{\bp_0}^p a' (u) \, du. 
    \]
    where the derivative $a'$ is integrable.
    Suppose furthermore that the rearrangement of $a$ satisfies
    \eqref{eq:sharp-cao}.  
    Evaluated at $p = \bp_0$, the pointwise bounds imply: 
    \begin{equation}
        a (\bp_0 ) = \by (\bp_0) = \psi_l (\bp_0) = \psi_u (\bp_0). 
        \label{eq:a-mean}
    \end{equation}
    Evaluated at $p \in [\bp_0, \bp_1]$, the (lower) bounds imply:  
    \begin{equation}
        I_{a' (U_c)} \left( \frac{p - \bp_0}{\bp_1 - \bp_0} \right)
        \geq 
        I_{\Delta_c^-} \left( \frac{p - \bp_0}{\bp_1 - \bp_0} \right)
        \quad \text{for $p \in [\bp_0, \bp_1]$} 
        \label{eq:a-ssd}
    \end{equation}
    This follows from plugging in $a (\bp_0) = \by (\bp_0)$ and recalling that: 
    \[
        \int_{\bp_0}^p (a')^{\uparrow_c} (u) \, du 
        =
        (\bp_1 - \bp_0) \cdot I_{a' (U_c)} 
        \left( \frac{p - \bp_0}{\bp_1 - \bp_0} \right)
        \quad \text{for $p \in [\bp_0, \bp_1]$.} 
    \]
    Finally, evaluated at $p = \bp_1$, the bounds imply:  
    \begin{equation}
        I_{a' (U_c)} (1) = \E [a' (U_c)] = \late_{IA}.  
        \label{eq:a-late}
    \end{equation}
    Combining \eqref{eq:a-ssd}, \eqref{eq:a-late}, and integrability implies
    that $a' \in \M^*$ by \autoref{thm:sharp-set}. 
    Combining $a' \in \M^*$ with \eqref{eq:a-mean} yields the sufficient
    representation \eqref{eq:a-rep}, and so $a \in \A^*$.  
    Finally, convexity follows from the sufficient representation
    \eqref{eq:a-rep} and convexity of the set $\M^*$.  

\end{proof}

\newpage

\section{Remaining Proofs (Online Appendix)}

\begin{proof}[Proof of \autoref{thm:group-lemma}]

    From the definition \eqref{eq:strata} and the latent index formulation
    \eqref{eq:selection}, 
    the groups are equivalently expressed in
    terms of unobservables: $\{ G = a \} = \{ U \leq \bp_0\}$, $\{ G = c \} = \{
    \bp_0 < U \leq \bp_1 \}$, and $\{ G = n \} = \{ \bp_1 < U \}$.  
    Therefore the group probabilities are given by $P(G=a) = \bp_0$, $P (G=c) =
    \bp_1 - \bp_0$, and $P(G=n) = 1 - \bp_1$, 
    and Bayes' rule implies that the conditional $U_g$ are uniformly distributed
    on their respective supports. 
    
    The observed distribution of $(D,Y,Z)$ is equivalently expressed as the
    probabilities $P(D=d,Z=z)$ and the conditional outcome cdfs $F_{Y | D=d,Z=z}
    = F_{Y_d | D=d,Z=z}$ for all $(d,z) \in \{0,1\}^2$.  
    Consider a partition of the sample space $\Omega$ into the observed sets $\{
    D = d, Z=z \}$. 
    Since $D = 1$ implies $G \in \{a,c\}$ and $D=0$ implies $G \in \{c,n\}$, a
    weakly finer partition than the sets $\{ D = d, Z=z \}$ is given by the sets
    $\{ G = g, Z = z\}$. 
    Furthermore, $\{D=1, Z = 0\} = \{D_0 = 1, Z=0\} = \{G=a, Z=0\}$. 
    Next, the earlier expression of $G$ as a function of $U$ combined with
    independence $U \perp Z$ imply that $G \perp Z$.
    Thus $\{G=a, Z=0\}$ is a random subset of $\{G=a\}$, for which $Y_1$ is
    observed.  
    Therefore $F_{Y_{1,a}} (y) = F_{Y | D=1,Z=0} (y )$. 
    By analogous reasoning, $F_{Y_{0,n}} (y) = F_{Y | D=0,Z=1} (y)$. 
    For compliers $G=c$, Imbens and Rubin~\cite{imbensrubin} show that the
    distributions of compliers outcomes $Y_{d,c}$ are identified
    for both treatment decisions $d \in \{0,1\}$; see also
    Abadie~\cite{abadie} for an expression of the complier outcome cdfs as a
    function of the data.

    Conversely, observe that within each group $\{G=g, Z=z\}$, at most one
    potential outcome $Y_d$ (or rather, its distribution) is observed, while
    any information on the other $Y_{1-d}$ is recovered by independence of $G$
    and $Z$. 
    Therefore the data and model impose no assumptions beyond consistency with
    the identified marginals on the conditional joint distribution of $Y_0$ and
    $Y_1$ within each set $\{G = g\}$.
    Since the partition into sets $\{U = u\}$ is finer than the partition into
    $\{G=g\}$, the result holds also for the conditional joint distribution of
    $(U,Y_0,Y_1)$ within each set $\{G=g\}$.  
    Thus any joint distribution consistent with the identified conditional
    marginals can be generated in a way that is consistent with the model and
    the data. 

\end{proof}

\begin{proof}[Proof of \autoref{thm:gini}]
    The proof follows from algebra and definitions: 
    \begin{align*}
        R (\bp_0 , \bp_1) 
        &= \int_{\bp_0}^{\bp_1} [ \psi_h (u) - \psi_l (u) ] \, du \\[0.05in]
        &= (\bp_1 - \bp_0) \int_0^1 \left[ 
            \psi_h (\bp_0 + (\bp_1 - \bp_0) q) - \psi_l (\bp_0 + (\bp_1 - \bp_0) q)
        \right] \, dq  \\
        &= (\bp_1 - \bp_0) \int_0^1 \left[
            \by (\bp_1) - \by (\bp_0) - (\bp_1 - \bp_0) 
            [ I_{\Delta_c^-} (1-q) + I_{\Delta_c^-} (q) ]
        \right] \, dq \\
        &= (\bp_1 - \bp_0)^2 \int_0^1 \left[ 
            \late_{IA} - 2 I_{\Delta_c^-} (q) 
        \right] \, dq \\
        &= (\bp_1 - \bp_0)^2 \cdot 2 \int_0^1 \left[ 
            q \cdot \late_{IA} - I_{\Delta_c^-} (q)
        \right] \, dq \\
        &= (\bp_1 - \bp_0)^2 \cdot \Gamma_{\Delta_c^-} 
    \end{align*}
\end{proof}

\begin{proof}[Proof of \autoref{thm:mte-nec-rank}]
    It suffices to show that $\mte (U_g) = \E [ \Delta_g^+ | U_g]$ for each $g
    \in \{a,c,n\}$, at which point necessity of the conditions follows   
    by the same reasoning as in the necessity proof of \autoref{thm:sharp-set}.  
    Observe:

    \vspace{-0.25in}
    \begin{align*}
        \mte (U_g) &= \E [ Y_{1,g} - Y_{0,g} | U_g ] \\  
            &= \E [ Q_{Y_1} (V_{1,g}) - Q_{Y_0} (V_{0,g}) | U_g ] \\
            &= \E [ Q_{Y_1} (V_{0,g}) - Q_{Y_0} (V_{0,g}) | U_g ] \\
            &= \E [ \Delta_g^+ | U_g ]
    \end{align*}
    where the first equality follows by definition of the $\mte$ function and
    the group-conditional random vector, the second equality by the Skorokhod
    representation implicit in the definition of rank similarity, the third
    equality by rank similarity, and the last equality by definition of
    $\Delta_g^+$.
\end{proof}

\begin{proof}[Proof of \autoref{thm:qq}] 
    The result is proved for realizations of $U$, from which it follows for any
    function thereof, i.e. random variables measurable with respect to $U$. 
    Let $Q Q_u$ denote the conditional $QQ$ plot for $U=u$.  
    By the Skorokhod quantile representation implicit in the assumption of rank
    similarity, the conditional quantile function of outcomes is equal to the
    unconditional quantile function of outcomes evaluated at the conditional
    quantile function of ranks. 
    That is, letting $Y_{d,u} \sim Y_d | U = u$ and $V_{d,u} \sim V_d | U = u$,     
    \[
        Q_{Y_{d,u}} (q) = Q_{Y_d} (Q_{V_{d,u}} (q))
        \quad \text{for all $q,u \in [0,1]$.}
    \]
    By rank similarity, the conditional distribution of rank variables are equal. 
    Expressed in terms of the quantile functions,   
    \[
        Q_{V_{0,u}} (q) = Q_{V_{1,u}} ( q )
        \quad \text{for all $q,u \in [0,1]$.}
    \]
   Combining these observations yields the desired result:  
   \begin{align*}
       Q Q_u
       & = \{ (Q_{Y_{0,u}} (q), Q_{Y_{1,u}} (q)): q \in [0,1] \} \\ 
       & = \{ Q_{Y_0} (Q_{V_{0,u}} (q)), Q_{Y_1} (Q_{V_{1,u}} (q)): q \in [0,1] \} \\
       & = \{ Q_{Y_0} (Q_{V_{0,u}} (q)), Q_{Y_1} (Q_{V_{0,u}} (q))
       : q \in [0,1] \} \subseteq Q Q. 
   \end{align*}
\end{proof}

\begin{proof}[Proof of \autoref{thm:extrapolate-quantiles}]
    \autoref{thm:qq} implies that $QQ_g \subseteq QQ$ for all $g \in \{a,c,n\}$. 
    \autoref{assn:support} implies that $QQ = QQ_c$, and so $QQ_g \subseteq QQ_c$. 
    Necessity of the bounds then follows by definition. 

    To prove uniform sharpness for $(d,g) \in \{ (0,a), (1,n) \}$, note first
    that the bounds are identified from the data because the quantile function
    $Q_{Y_{1-d,g}}$ is identified in the basic model
    (\autoref{assn:data} and \autoref{assn:lism}). 
    It remains to show that the bounds are attained by a data-generating process
    consistent with the assumptions. 
    For any bound, consider a rank-invariant Skorokhod representation, e.g.
    $(Y_{0,n}, Y_{1,n}) \stackrel{d}{=} (Q_{Y_{0,n}} (V),
    \underline{Q}_{Y_{1,n}} (V))$ for $V \sim U[0,1]$.  
    This satisfies the basic model, which imposed no restrictions on the
    extrapolated quantile function.  
    Additionally, its quantile-quantile plot belongs to the true population plot
    $QQ$ by construction of the extrapolated quantile function,
    \autoref{thm:qq}, and the closed complete support condition of
    \autoref{assn:support}. 
    This guarantees that the rank invariant construction by group $g$ is also
    rank-invariant in the population, and thus satisfies (either version of)
    \autoref{assn:rank}. 
\end{proof}

\begin{proof}[Proof of \autoref{thm:sharp-te-rank}] 
    As in the proof of \autoref{thm:sharp-te}, fix $w$, consider an
    $\overline{m}_w$ satisfying \eqref{eq:te-sharp-rank}, and for consistency of
    notation let $ M_g = m (U_g)$, $W_g = w (U_g)$, $\bar{M}_g = \overline{m}_w
    (U_g)$, and $V \sim U[0,1]$. 

    Suppose momentarily that the distribution of each $\Delta_g^+$ were
    identified. 
    In that case, following the logic of \autoref{thm:sharp-set}, the necessary
    characterization of \autoref{thm:mte-nec-rank} becomes sufficient for
    characterizing the set $\M^+$ of empirically consistent $\mte$ functions
    under the additional \autoref{assn:rank} and \autoref{assn:support}. 
    Then the desired result follows analogously to the proof of
    \autoref{thm:sharp-te} upon decomposing the functional into its group
    conditional components: 
    \[
        \int_0^1 m(u) w(u) \, du = \sum_g \E [ M_g W_g ] \cdot P(G=g), 
    \]
    and maximizing piecewise. 
    Namely, analogous reasoning to the proof of \autoref{thm:sharp-te} yields: 
    \[
        \int_0^1 m(u) w(u) \, du \leq 
        \sum_g \E [ Q_{\Delta_g^+} (V) Q_{W_g} (V)] \cdot P (G = g)
    \]

    Now consider the case where the distribution of $\Delta_g^+$ is not fully
    identified, but rather bounded by \autoref{thm:rank-te-fsd}, i.e. $\Delta_g^+
    \preceq_{FSD} \overline{\Delta}_g^+$. 
    By assumption, $w (u) \geq 0$ for all $u$, so that also $Q_{W_g} (V) \geq 0$. 
    Combining these observations, it follows that: 
    \[
        \E [ Q_{\Delta_g^+} (V) Q_{W_g} (V)] \leq 
        \E [ Q_{\overline{\Delta}_g^+} (V) Q_{W_g} (V)]
    \]
    for each $g$. 
    By definition of $\overline{m}_w$, the right hand side is equal to $\E [
    \overline{M}_g W_g]$ for each $g$, and so:  
    \[
        \int_0^1 m (u) w (u) \, du \leq \int_0^1 \overline{m}_w (u) w (u) \, du
    \]
    for all $m \in \M^+$.  
    Finally, the underlying group-conditional treatment effects
    $\overline{\Delta}_g^+$ are jointly attainable by \autoref{thm:rank-te-fsd},
    which proves the sharpness of the upper bound.
    An argument for the lower bound follows analogously. 
\end{proof}

\begin{proof}[Proof of \autoref{thm:bounds-cao-rank}] 
    First, it will be useful to observe that the ``reverse'' integrated quantile
    function, which aggregates quantiles in decreasing order, can be expressed
    as:  
    \begin{equation}
        \label{eq:iqf-rev}
        \int_q^1 Q_X (v) \, dv = \E [ X ] - I_X (q)
    \end{equation} 
    for any integrable $X$ and $q \in [0,1]$ because $\int_0^1 Q_X (v) \, dv =
    \E [ X ]$. 

    Combining \autoref{thm:mte-nec-rank} and
    \autoref{thm:rank-te-fsd} implies: 
    \begin{equation}
        \label{eq:outer-bound1}
        I_{\mte (U_g)} (q) \geq 
        I_{\Delta_g^+} (q) \geq
        I_{\underline{\Delta}_g^+} (q) 
    \end{equation}
    for all $q \in [0,1]$, by definition of SSD and the fact that FSD is a
    stronger ordering (i.e. FSD implies SSD).  
    Additionally, the combination of aforementioned results implies: 
    \begin{equation}
        \label{eq:outer-bound2}
        \E [ \mte (U_g) ] - I_{\mte (U_g)} (q) 
        \leq 
        \E [ \Delta_g^+ ] - I_{\Delta_g^+} (q) \leq 
        \E [ \overline{\Delta}_g^+ ] - I_{\overline{\Delta}_g^+} (q) 
    \end{equation}
    The first inequality follows from equality of means and SSD; by
    \eqref{eq:iqf-rev}, the second inequality is equivalent to:     
    \[
        \int_q^1 Q_{\Delta_g^+} (v) \, dv \leq \int_q^1
        Q_{\overline{\Delta}_g^+} (v) \, dv
    \]
    which follows from FSD. 

    Now consider parameter values among the always-treated, i.e. $p \in [0,
    \bp_0]$. 
    By virtue of increasing and decreasing rearrangements it must be that: 
    \begin{align*}
        \int_p^{\bp_0} Q_{\mte (U_a)} \left( 
            \frac{\bp_0 - u}{\bp_0}
        \right) \, du 
        \leq 
        \int_{p}^{\bp_0} \mte (u) \, du
        \leq
        \int_p^{\bp_0} Q_{\mte (U_a)} \left( 
            \frac{u}{\bp_0} 
        \right) \, du
    \end{align*}
    Equivalently, by a change of variables and invoking \eqref{eq:iqf-rev},  
    \begin{align*}
        \bp_0 \cdot I_{\mte (U_a)} \left( \frac{\bp_0 - p}{\bp_0} \right)
        \leq 
        \int_{p}^{\bp_0} \mte (u) \, du
        & \leq
        \bp_0 \cdot \left[ 
            \E [ \mte (U_a) ] - I_{\mte (U_a)} \left( \frac{p}{\bp_0} \right)
        \right]
    \end{align*}
    Then invoking the inequalities \eqref{eq:outer-bound1} and
    \eqref{eq:outer-bound2} implies: 
    \begin{align*}
        \bp_0 \cdot I_{ \underline{\Delta}_a^+ } \left( \frac{\bp_0 - p}{\bp_0} \right)
        \leq 
        \int_{p}^{\bp_0} \mte (u) \, du
        & \leq
        \bp_0 \cdot \left[ 
            \E [ \overline{\Delta}_a^+ ] - I_{\overline{\Delta}_a^+} \left( \frac{p}{\bp_0} \right)
        \right]
    \end{align*}
    Combining these bounds with the fact that $\by (p) = \by (\bp_0) -
    \int_{p}^{\bp_0} \mte (u) \, du$ then yields the desired result among the
    always-treated, $p \in [0,\bp_0]$. 
    Analogous reasoning yields the bounds among the never-treated, $p \in
    [\bp_1, 1]$. 
    The complier bounds are similarly obtained but invoke the stronger
    inequalities in \eqref{eq:outer-bound1} and \eqref{eq:outer-bound2}. 
    Finally, uniform sharpness of these bounds follows from
    \autoref{thm:rank-te-fsd}.  

\end{proof}

\begin{proof}[Proof of \autoref{thm:cond-rs}]
    To rank the assumptions, suppose the strong version
    (\autoref{assn:rank-cov}b) holds, and let $V_{d,x} \sim (V_d | X=x)$ and
    $\tilde{V}_d = F_{V_{d,X}} (V_d)$.  
    Such $\tilde{V}_d$ are identically distributed conditional on $(X,U)=(x,u)$
    because $V_0$ and $V_1$ are conditionally identically distributed, which
    further implies $F_{V_{0,x}} (\cdot) = F_{V_{1,x}} (\cdot)$.
    Additionally, using the Skorokhod representation's implication that 
    $Q_{Y_{d,x}} (q) = Q_{Y_d} (Q_{V_{d,x}} (q))$, it follows that:   
    \[
        Q_{Y_{d,X}} (\tilde{V}_d) = Q_{Y_d} (Q_{V_{d,X}} (F_{V_{d,X}} (V_d))) 
        \stackrel{a.s.}{=} Q_{Y_d} (V_d) = Y_d. 
    \]
    Thus $\tilde{V}_d$ satisfy \autoref{assn:rank-cov}a (almost surely). 
    The first testable implication \eqref{eq:strong1} follows immediately from
    \autoref{thm:qq}.
        For the second implication, first note that while the rank random variables
    $V_d$ or $\tilde{V}_d$ need not be unique if potential outcomes are not
    continuous, any possible rank variables induce the same outcomes. 
    Therefore the implication follows immediately from the definition of the
    respective comonotonic effects and the fact that the strong rank invariance
    assumption implies the weak version. 
\end{proof}

\section{Relation to Distribution of Treatment Effects (Online Appendix)}
\label{apx:te-example}

This appendix provides a pair of counter-examples illustrating that i) neither
the second-order bounds \eqref{eq:te-ssd} nor ii) the combination of first-order
Makarov bounds and second-order bounds characterize the sharp set of treatment
effects.
In addition, this appendix provides a simple example where the $\mte$ relaxation
of the quantile functional bounding problem provides different but still
intuitive answers.  

First, recall the Makarov~\cite{makarov} bounds on the sum of two random
variables with known marginal distributions, introduced into the treatment
effect framework by Fan and Park~\cite{fanpark2010}.  
Respectively let $\Delta^L$ and $\Delta^U$ be random variables with the
following distributions:  
\begin{align*}
    F_{\Delta^L} (\delta) &= \sup_y \max \{ F_{Y_1} (y) - F_{Y_0} (y - \delta), 0 \} \\
    F_{\Delta^U} (\delta) &= 1 + \inf_y \min \{ F_{Y_1} (y) - F_{Y_0} (y -
    \delta), 0 \}
\end{align*}
The Makarov bounds are summarized by the first-order stochastic dominance
relation: 
\begin{equation}
    \Delta^L \preceq_{FSD} \Delta \preceq_{FSD} \Delta^U
    \label{eq:te-fsd}
\end{equation}
As observed by Fan and Park~\cite{fanpark2010}, \eqref{eq:te-fsd} provides 
\emph{pointwise} sharp bounds on the distribution and quantile functions of the
treatment effect $\Delta$.%
\footnote{
    In contrast, by the integrated quantile/distribution representations of SSD and
    feasibility of the bounding random variables, the second-order relation
    \eqref{eq:te-ssd} provides \emph{uniformly} sharp bounds on the
    \emph{integrated} quantile and distribution functions.
    To the best of my knowledge, this is a new (if minor) observation in the
    literature.  
}

Neither the first-order stochastic relation \eqref{eq:te-fsd} nor the
second-order stochastic relation \eqref{eq:te-ssd} characterizes the set of
possible treatment effects.
The first observation is due to Firpo and Ridder~\cite{firporidder2019}, and the
second observation is shown by counter-example.
A fortiori, I show (again by counter-example) that the combination of
first and second-order relations \eqref{eq:te-fsd} and \eqref{eq:te-ssd} does
not characterize the sharp set of treatment effect distributions.%
\footnote{
    It is worth noting that the set of distributions satisfying both
    \eqref{eq:te-fsd} and \eqref{eq:te-ssd} is strictly sharper than the set of
    distributions studied by Firpo and Ridder~\cite{firporidder2019}, since it
    imposes requirements on moments of the distribution beyond the mean. 
}

For the pair of counter-examples, it suffices to consider the case where $P (Y_d
= y) = 1/2$ for $(d,y) \in \{0,1\}^2$. 
In that case, every realized treatment effect belongs to the set $\{ -1, 0,
1\}$. 
Thus feasible distribution functions $F$ are summarized by their values at
three points, $(F (-1), F(0), F(1))$. 
In fact, the same is true of the distribution functions corresponding to the
first-order Makarov bounds $F_{\Delta^L} = (0,1/2,1)$ and $F_{\Delta^U} =
(1/2,1,1)$, as well as the second-order bounds $F_{\Delta^-} = (1/2,1/2,1)$ and
$F_{\Delta^+} = (0,1,1)$. 
For the second-order bounds, the integrated distribution functions (IDF)
$\tilde{I} (\cdot)$ are then given by: 
\begin{equation}               
    \label{eq:ex-idfs} 
    \tilde{I}_{\Delta^+} (\delta) = \begin{cases} 
        0 & \text{for $\delta \in [-1,0)$} \\
        \delta & \text{for $\delta \in [0,1]$} 
    \end{cases} 
    \quad 
    \text{and} 
    \quad
    \tilde{I}_{\Delta^-} (\delta) = \frac{\delta + 1}{2} 
    \text{ for $\delta \in [-1,1]$.}
\end{equation}
For the first counter-example, consider: 
\[
    F_1 (\delta) = \begin{cases} 
        \frac{1}{3} & \text{for $\delta \in \left[ -1, \frac{1}{2} \right)$}  \\
        1 & \text{for $\delta \in \left[ \frac{1}{2},1 \right]$} 
    \end{cases} 
    \quad \text{with} \quad
    \tilde{I}_1 (\delta) = \begin{cases}
        \frac{\delta+1}{3} & \text{for $\delta \in \left[ -1,\frac{1}{2} \right)$} \\
        \delta & \text{for $\delta \in \left[ \frac{1}{2},1 \right]$} 
    \end{cases}
\]
This distribution satisfies the second-order relation \eqref{eq:te-ssd} because
the IDF is pointwise bounded by the extremal IDFs in \eqref{eq:ex-idfs}.  
However, it does not satisfy the necessary first-order Makarov bounds
\eqref{eq:te-fsd}. 
For the second counter-example, consider: 
\[
    F_2 (\delta) = \begin{cases}
        0 & \text{for $\delta \in \left[ -1,-\frac{1}{3} \right)$} \\
        \frac{1}{2} & \text{for $\delta \in \left[ - \frac{1}{3},\frac{1}{3} \right)$} \\
        1 & \text{for $\delta \in \left[ \frac{1}{3},1 \right]$}
    \end{cases} 
    \quad \text{with} \quad
    \tilde{I}_2 (\delta) = \begin{cases} 
        0 & \text{for $\delta \in \left[ -1, \frac{1}{3} \right)$} \\
        \frac{1}{2} \left[ \delta + \frac{1}{3} \right] & 
        \text{for $\delta \in \left[- \frac{1}{3}, \frac{1}{3} \right)$}  \\
        \delta & \text{for $\delta \in \left[ \frac{1}{3},1 \right]$}
    \end{cases}
\]
This distribution satisfies both the first-order and second-order conditions
\eqref{eq:te-fsd} and \eqref{eq:te-ssd}, but its mass points are not in the set
$\{ -1, 0, 1\}$. 
Therefore it is not a feasible treatment effect distribution. 

For the final example, recall from \eqref{eq:gini} that the Gini coefficient has
a simple expression in terms of the quantile function in the case of a nonzero
mean. 
Expressed as a functional $\Phi (\cdot)$ of the quantile function $Q_\Delta$, 
\begin{equation}
    \Phi ( Q_\Delta ) = 1 - \frac{2}{\E [\Delta]} \int_0^1 \int_0^q Q_\Delta (u) \, du \, dq  
    \label{eq:gini-functional}
\end{equation}
Since the Gini coefficient is a spread parameter in the sense of
Stoye~\cite{stoye}, it follows by observation of Fan and Park~\cite{fanpark2010}
that the Gini coefficient of the treatment effect is bounded above and below by
(and attains) its values in the extreme cases where the treatment effect is
respectively perfectly counter- and co-monotonic. 
Now consider the relaxed problem of maximizing and minimizing the Gini
functional $\Phi (\cdot)$ over feasible $\mte$ functions $m \in \M^*$.  
The relaxed maximization problem still recovers the countermonotonic upper
bound;  in contrast, the relaxed lower bound is equal to zero, a value attained
when marginal treatment effects are homogeneous.  
Typically this is strictly lower than the lower bound over quantile functions.%
\footnote{
    Any pair of marginal distributions is consistent with a constant
    \emph{marginal} treatment effect, but many such pairs are inconsistent with a
    constant treatment effect.
    For example, take $P(Y_0 = y) = 1/2$ for $y = 0,1$ and $P(Y_1 = 1) = 1$.   
}
Nevertheless, the lower bound of zero intuitively captures the minimal amount of
ex ante dispersion --- that is, dispersion in individuals' expected (rather than
realized) treatment effects.  
Thus the relaxed lower bound captures an interesting 
feature, despite differing from the original bound of interest. 

\end{document}

%% file: empirical/tables/wte-bounds.tex
\begin{table}[H]
    \centering

    \setlength{\tabcolsep}{1.5em} 

\resizebox{\textwidth}{!}{%
    \begin{tabular}{@{}l@{\qquad}cccccc@{}} \toprule
        \multicolumn{7}{c}{Model}                                                      \\ \cmidrule(r){2-7} 
        \multirow{-2}{*}{\makecell[l]{Parameter}} & Means & Base & Covariate & Conditional Rank Sim. & Rank Sim. & Linearity \\ \midrule
        \multicolumn{7}{c}{ \small Panel A. Outcome: Visited ER, Covariate: Visited ER in Pre-Period, $N =$ 19,643}  \\[0.05in]
        $\late(0,\bp_0)$                                & $[-0.45, 0.55]$ &  $[-0.45, 0.55]$ & $[-0.45, 0.55]$ & $[0,    0.55]$ & $[0,    0.55]$ & $ \{ 0.13  \}$ \\ 
        $\late(\bp_0,\bp_1)$                            & $ \{ 0.05 \} $  &  $ \{ 0.05 \} $  & $ \{ 0.05 \} $  & $ \{ 0.05 \} $ & $ \{ 0.05 \} $ & $ \{ 0.05  \}$ \\ 
        $\late(\bp_1, 1)$                               & $[-0.31, 0.69]$ &  $[-0.31, 0.69]$ & $[-0.31, 0.69]$ & $[0,    0.69]$ & $[0,    0.69]$ & $ \{ -0.10 \}$ \\ 
        $\late(\bp_0, \bp_1 - 0.1(\bp_1 - \bp_0))$      & $[-0.05, 0.17]$ &  $[-0.05, 0.17]$ & $[-0.06, 0.17]$ & $[0,    0.06]$ & $[0,    0.06]$ & $ \{ 0.06  \}$ \\ 
        $\late(\bp_0, \bp_1 + 0.1(\bp_1 - \bp_0))$      & $[-0.04, 0.14]$ &  $[-0.04, 0.14]$ & $[-0.05, 0.14]$ & $[0.05, 0.14]$ & $[0.05, 0.14]$ & $ \{ 0.05  \}$ \\ 
        $\late(0, 1)$                                   & $[-0.24, 0.50]$ &  $[-0.24, 0.50]$ & $[-0.24, 0.50]$ & $[0.01, 0.50]$ & $[0.01, 0.50]$ & $ \{ -0.02  \}$\\ \midrule 
        \multicolumn{7}{c}{ \small Panel B. Outcome: Number of ER Visits, Covariate: Number of ER Visits in Pre-Period (Truncated at 10), $N =$ 19,615 }  \\[0.05in]
        $\late(0,\bp_0)$                                & $(-\infty, 1.88]$   & $(-\infty, 1.88]$  & $(-\infty, 1.88]$   & $[-0.18, 0.44]$ & $[0.12, 0.63]$ & $\{ 0.55  \}$ \\ 
        $\late(\bp_0,\bp_1)$                            & $\{ 0.27     \} $   & $\{ 0.27     \} $  & $\{ 0.28     \} $   & $ \{ 0.28   \}$ & $ \{ 0.27  \}$ & $\{ 0.27  \}$ \\ 
        $\late(\bp_1, 1)$                               & $[-0.85, \infty)$   & $[-0.85, \infty)$  & $[-0.85, \infty)$   & $[0.06,  1.17]$ & $[0.07, 1.08]$ & $\{ -0.31 \}$ \\ 
        $\late(\bp_0, \bp_1 - 0.1(\bp_1 - \bp_0))$      & $[-1.30,   1.60]$   & $[-0.66,   1.11]$  & $[-0.53,   0.98]$   & $[0.00,  0.43]$ & $[0.13, 0.30]$ & $\{ 0.29  \}$ \\ 
        $\late(\bp_0, \bp_1 + 0.1(\bp_1 - \bp_0))$      & $[-1.51, \infty)$   & $[-0.56, \infty)$  & $[-0.56, \infty)$   & $[0.06,  0.79]$ & $[0.25, 0.51]$ & $\{ 0.25  \}$ \\ 
        $\late(0, 1)$                                   & $(-\infty, \infty)$ & $(-\infty, \infty)$ & $(-\infty, \infty)$ & $[0.08,  0.83]$ & $[0.13, 0.80]$ & $\{ -0.03 \}$ \\ \midrule 
        \multicolumn{7}{c}{ \small Panel C. Outcome: ER Total Charges, Covariate: Upper Quintile of ER Total Charges in Pre-Period, $N =$ 19,619 }  \\[0.05in]
        $\late(0,\bp_0)$                                & $(-\infty, 8777]$   & $(-\infty, 8777]$   & $(-\infty, 8773]$   & $[813, 1274]$ & $[1148, 1517]$ & $ \{ 5082  \}$ \\ 
        $\late(\bp_0,\bp_1)$                            & $ \{ 516     \} $   & $ \{ 436     \} $   & $ \{ 466     \} $   & $ \{ 466 \} $ & $ \{ 437  \} $ & $ \{ 516   \}$ \\ 
        $\late(\bp_1, 1)$                               & $[-2929, \infty)$   & $[-2932, \infty)$   & $[-2933, \infty)$   & $[598, 1153]$ & $[562,  1092]$ & $ \{ -8916 \}$ \\ 
        $\late(\bp_0, \bp_1 - 0.1(\bp_1 - \bp_0))$      & $[-3828,   4402]$   & $[-2981,   3523]$   & $[-2948,   3531]$   & $[-152, 721]$ & $[-100,  651]$ & $ \{ 804   \}$ \\ 
        $\late(\bp_0, \bp_1 + 0.1(\bp_1 - \bp_0))$      & $[-5579, \infty)$   & $[-3824, \infty)$   & $[-3797, \infty)$   & $[223, 2036]$ & $[217,  1831]$ & $ \{ 228   \}$ \\ 
        $\late(0, 1)$                                   & $(-\infty, \infty)$  & $(-\infty, \infty)$ & $(-\infty, \infty)$ & $[596,  993]$ & $[618,   986]$ & $ \{ -4350 \}$ \\ \bottomrule 
    \end{tabular}
}
    
\caption{
    \label{tab:lates} 
    Sharp Bounds for Various Counterfactual $\late$ Parameters, OHIE Data. 
    This table provides point estimates of sharp bounds
    for selected counterfactual $\late$ parameters discussed in the
    text across different model specifications for three ER utilization
    outcomes: whether a participant visited the ER, the number of ER
    visits, and the total ER charges incurred.
    Each model is estimated on all observations with non-missing
    outcome and covariate values.   
    Point-identifying bounds are denoted in brackets.  
    Finally, some discrepancies in point estimates 
    arise because the
    complier means in the distribution models are derived from
    estimates of the entire complier outcome distributions; 
    in the case of the highly dispersed total charges outcomes, this leads the
    mean estimate to be highly dependent on the estimate of the
    distributions' extreme right tails. 
}

\end{table}